\documentclass[12pt]{article}
\usepackage{amsfonts,amsthm}
\usepackage{amsmath,amssymb}
\usepackage{array}
\usepackage{cite}
\usepackage{mathrsfs}
\usepackage{dsfont}
\usepackage[title]{appendix}

\usepackage{footmisc}
\usepackage{axodraw2}
\usepackage{caption}
\usepackage{graphicx}
\usepackage{comment}
\usepackage{slashed}
\usepackage{cancel}
\usepackage{booktabs}
\usepackage{subcaption}
\usepackage{tcolorbox}
\usepackage{graphicx}
\usepackage{tikz}
\usetikzlibrary{decorations.pathmorphing,decorations.markings}
\tikzset{
every node/.style={font=\footnotesize},
photon/.style={decorate, decoration={snake,amplitude=2pt, segment length=5pt}, draw=black},
particle/.style={draw=black, postaction={decorate}, decoration={markings,mark=at position .5 with {\arrow[draw=black]{>}}}},
antiparticle/.style={draw=black, postaction={decorate}, decoration={markings,mark=at position .5 with {\arrow[draw=black]{>}}}},
gluon/.style={decorate, draw=black, decoration={coil,amplitude=4pt, segment length=5pt}}
goldstone/.style={draw=green,postaction={decorate},decoration={markings,mark=at position .5 with {\arrow[draw=blue]{>}}}}
}
\usepackage[
top    = 3.5cm,
bottom = 3.5cm,
left   = 3.5cm,
right  = 3.5cm]{geometry}
\usepackage[hypertexnames=false,pdfencoding=auto]{hyperref}
\usepackage[all]{hypcap}        % needed to help hyperlinks direct correctly;

\newcommand*\circled[1]{%
  \tikz[baseline=(char.base)]{
    \node[circle,draw,minimum size=1.2em,inner sep=0pt,font=\footnotesize]
    (char) {#1};
  }%
}

\newenvironment{Eqnarray}%
     {\arraycolsep 0.14em\begin{eqnarray}}{\end{eqnarray}}
\newcommand{\ben}{\begin{enumerate}}
\newcommand{\een}{\end{enumerate}}
\newcommand{\be}{\begin{equation}}
\newcommand{\ee}{\end{equation}}
\newcommand{\bea}{\begin{Eqnarray}}
\newcommand{\eea}{\end{Eqnarray}}

\newcommand{\doublet}[2]{ \left(\begin{array}{c}#1 \\[0.5ex] #2\end{array}\right) }

\newcommand{\Z}{\mathbb{Z}}

\newcommand\Tstrut{\rule{0pt}{2.9ex}}         % "top" strut
\newcommand\Bstrut{\rule[-1.2ex]{0pt}{0pt}}   % "bottom" strut

\usepackage{color}

\def\lsim{\mathrel{\rlap{\lower4pt\hbox{\hskip1pt$\sim$}}
    \raise1pt\hbox{$<$}}}
\def\gsim{\mathrel{\rlap{\lower4pt\hbox{\hskip1pt$\sim$}}
    \raise1pt\hbox{$>$}}}

\def\O{ {\cal O }}

\def\<{\left\langle}
\def\>{\right\rangle}

\newcommand{\bt}{\begin{tabular}}
\newcommand{\et}{\end{tabular}}

\def\wtil{\widetilde}

\def\ifmath#1{\relax\ifmmode #1\else $#1$\fi}

\def\L{ {\cal L }}

\def\mzd{m_Z^2}

\def\ls#1{\ifmath{_{\lower1.5pt\hbox{$\scriptstyle #1$}}}}
\def\lss#1{\ifmath{^{\,\lower2.5pt\hbox{$\scriptstyle #1$}}}}
\def\lsup#1{^{\lower 6pt\hbox{$\scriptstyle#1$}}}
\def\llsup#1{^{\lower 3pt\hbox{$\scriptstyle#1$}}}
\def\lasup#1{^{\lower 2pt\hbox{$\scriptstyle#1$}}}
\def\nicefrac#1#2{\hbox{$\frac{#1}{#2}$}}

\def\ddel{\!\!\mathrel{\raise1.5ex\hbox{$\leftrightarrow$\kern-.85em
\lower1.7ex\hbox{$\partial$}}}}

\def\thefootnote{\fnsymbol{footnote}}

\def\lsim{\mathrel{\rlap{\lower4pt\hbox{\hskip1pt$\sim$}}
    \raise1pt\hbox{$<$}}}         %less than or approx. symbol
\def\gsim{\mathrel{\rlap{\lower4pt\hbox{\hskip1pt$\sim$}}
    \raise1pt\hbox{$>$}}}         %greater than or approx. symbol

\newcommand{\bft}{\begin{footnotesize}}
\newcommand{\eft}{\end{footnotesize}}
\newcommand{\bts}{\begin{tiny}}
\newcommand{\ets}{\end{tiny}}

\usepackage[compat=1.0.0]{tikz-feynman}
 \tikzfeynmanset{compat=1.0.0}
\usepackage{tikz}
\usetikzlibrary{decorations.pathmorphing,decorations.markings}

\tikzset{
photon/.style={decorate, decoration={snake,amplitude=2pt, segment length=5pt}, draw=black},
particle/.style={draw=black, postaction={decorate}, decoration={markings,mark=at position .5 with {\arrow[draw=black]{>}}}},
antiparticle/.style={draw=black, postaction={decorate}, decoration={markings,mark=at position .5 with {\arrowreversed[draw=black]{>}}}},
gluon/.style={decorate, draw=black, decoration={coil,amplitude=4pt, segment length=5pt}},
goldstone/.style={draw=green,postaction={decorate},decoration={markings,mark=at position .5 with {\arrow[draw=blue]{>}}}}
}

\usepackage{color}
\usepackage{xcolor}
\definecolor{grey}{cmyk}{0,0,0,0.75}
\definecolor{tangerine}{cmyk}{0,0.5,1,0}
\definecolor{darkgreen}{cmyk}{1,0,1,0.23}
\definecolor{Red}{rgb}{1,0,0}
\definecolor{Blue}{rgb}{0,0,1}
\definecolor{Green}{rgb}{0,1,0}
\definecolor{Grey}{cmyk}{0,0,0,0.75}
\definecolor{Tangerine}{cmyk}{0,0.5,1,0}
\definecolor{Darkgreen}{cmyk}{1,0,1,0.23}
\definecolor{Cyan}{cmyk}{1,0,0,0}
\definecolor{Yellow}{cmyk}{0,0,1,0}
\allowdisplaybreaks[2]
\begin{document}

\begin{flushright}
\normalsize{
\phantom{x}\\[-50pt]
DIAS-STP-26-15 \\
CERN-TH-2026-237 \\
\today
}
\end{flushright}
\vspace{0.5cm}

\begin{center}
\LARGE\bf\boldmath
%\vspace*{2cm}
%A new class of CP violation structurally invisible to EDMs
%New CP-violating sources with zero EDM
New CP-violating sources beyond the reach of EDMs

\unboldmath
\end{center}
\vspace{0.4cm}
\begin{center}

Venus~Keus\footnote{E-mail:{\tt venus@stp.dias.ie, venus@cern.ch}} \\

\vspace{0.4cm}
{\sl Theoretical Physics Department, CERN, 1 Esplanade des Particules,\\
Geneva 23, CH-1211, Switzerland}\\[0.4cm]
{\sl Dublin Institute for Advanced Studies, School of Theoretical Physics,\\
10 Burlington road, Dublin, D04 C932, Ireland}\\[0.4cm]
{\sl Department of Physics and Helsinki Institute of Physics,\\
University of Helsinki, Gustaf H\"allstr\"omin katu 2,  Helsinki, FIN-00014, Finland}\\[0.2cm]
\end{center}
\vspace{0.2cm}

\renewcommand{\thefootnote}{\arabic{footnote}}
\setcounter{footnote}{0}

\begin{abstract}
The Standard Model does not contain sufficient amount of CP violation to account for the observed baryon asymmetry of the universe. Beyond Standard Model scenarios that aim to explain this asymmetry introduce new sources of CP violation, but these generically feed into electric dipole moments (EDMs), which are severely constrained by experiment. We indentify a new class of CP-violating sources that cannot generate EDMs directly: EDM operators are C conserving and P violating, whereas CP violation introduced here is, by construction, C violating and P conserving. We give a concrete model realisation accommodating such sources and
discuss their relevance for baryogenesis.
\end{abstract}

\vspace{15mm}

{\scriptsize Dedicated to the memory of Mark A. Rayner, a brilliant physicist, an exceptional science communicator, and a dear friend.}

\newpage

\setcounter{page}{1}

%%%%%%%%%%%%%%%%%%%%%%%%%%%%%%%%%%%%
%%%%%%%%%%%%%%%%%%%%%%%%%%%%%%%%%%%%
\section{Introduction}
\label{sec:intro}
%%%%%%%%%%%%%%%%%%%%%%%%%%%%%%%%%%%%
%%%%%%%%%%%%%%%%%%%%%%%%%%%%%%%%%%%%

%Searches for permanent electric dipole moments (EDMs) of elementary particles, nucleons, nuclei, atoms, and molecules provide some of the most sensitive probes of CP violation beyond the Standard Model (SM). While the SM contains two sources of CP violation, namely the complex phase of the Cabibbo-Kobayashi-Maskawa (CKM) matrix and the QCD $\bar{\theta}$ term, neither is expected to generate observable EDMs at current experimental sensitivities. The contribution from CKM-induced CP violation is highly suppressed by flavour-changing loop effects, whereas the non-observation of EDMs constrains the QCD vacuum angle to be extraordinarily small, giving rise to the strong CP problem.

CP violation is one of the necessary ingredients, as shown by Sakharov~\cite{Sakharov:1967dj}, of any successful baryogenesis mechanism to explain the observed matter-antimatter asymmetry of the Universe.
The Standard Model (SM) contains two sources of CP violation, namely the complex phase of the Cabibbo-Kobayashi-Maskawa (CKM) matrix and the QCD $\bar{\theta}$ term.
However, the CP violation present in the SM is insufficient to account for the observed baryon asymmetry~\cite{Gavela:1993ts,Gavela:1994dt,Huet:1994jb}.

This motivates frameworks beyond the SM proposing new sources of CP violation. Such sources, however, generically induce electric dipole moments (EDMs) that are tightly constrained by present experiments and will be probed even more stringently by the next generation of measurements.
%Consequently, EDM experiments provide a powerful and complementary window into new physics, probing
%flavour-diagonal
%CP-violating interactions at energy scales that often exceed the direct reach of collider experiments.
However, there is an often-neglected caveat; that is, the observation of EDMs would constitute direct evidence of parity (P) and time-reversal (T) violation and, assuming CPT invariance, of CP violation.  Correspondingly, EDMs arise from operators that are P violating and C conserving.

In this paper, we identify a new class of CP-violating sources that arises from C-violating and P-conserving interactions, and hence is structurally distinct from the C-conserving, P-violating interactions probed by EDM searches, despite being CP violating. Such new sources emerge from models featuring an extended dark sector composed entirely of scalar fields.

The key observation is that CP violation originating purely in the scalar sector is necessarily C violating and P conserving: parity is automatically conserved in any renormalisable theory of scalars and (electroweak) gauge bosons in the absence of Yukawa couplings. We demonstrate this general principle in a concrete framework of the $\mathbb{Z}_2$-symmetric 3-Higgs-doublet model (3HDM) with two \textit{inert} and one \textit{active} Higgs doublets, also known as I(2+1)HDM, in which explicit CP violation is confined to the dark (inert) sector, hence referred to as \textit{dark CP violation}.

This class of CP violation was first introduced in Ref.~\cite{Cordero-Cid:2016krd} and subsequently studied in Refs.~\cite{Cordero-Cid:2018man,Keus:2019szx,Keus:2020ooy,Keus:2021dti,Keus:2021ojv,Keus:2021nff,Keus:2024khd,Keus:2025ova}\footnote{It was shown that dark CP violation provides a handle on the otherwise fixed gauge-scalar couplings and opens up a large region of the parameter space to accommodate a CP-violating Dark Matter candidate in agreement with cosmological and collider experiments~\cite{Keus:2019szx}.}. It was previously understood that the corresponding contributions to EDMs are severely suppressed, arising only at three-loop order or beyond.

In the present work, we argue that the implications may be considerably stronger: the C-violating, P-conserving nature of this class of CP violation suggests a symmetry-based mechanism for evading EDM constraints, rather than relying on numerical suppression or accidental cancellations between individual diagrams.
An essential ingredient is the exact $\mathbb{Z}_2$ symmetry, which forbids Yukawa couplings of the inert scalars and prevents the direct transmission of dark CP violation to the fermionic sector.\footnote{This is in contrast to the CP-violating 2HDMs in which CP violation can enter the fermionic sector and generate non-zero EDM contributions, as discussed below.}

Although this class of CP violation does not directly generate EDM operators, it offers distinct experimental signatures, particularly through triple-gauge couplings (TGCs). A potential challenge to its EDM invisibility arises when these vertices are embedded into higher-order diagrams involving the parity-violating fermionic interactions of the SM. While the C-odd, P-even TGC structures cannot generate the C-even, P-odd EDM operator through C- and P-conserving interactions, the axial gauge-fermion couplings of the SM may, in principle, be able to circumvent this restriction. We nevertheless conjecture that the absence of EDM contributions from this type of CP violation persists in the full electroweak theory, potentially to all orders in perturbation theory.\footnote{Interestingly, this conjecture is consistent with the very recent explicit three-loop calculation of Ref.~\cite{Banno:2026hsc}, in which contributions associated with the $ZZ\gamma$ and $\gamma\gamma\gamma$ vertices were found to vanish through cancellations between opposite orientations of the internal fermion loop, whereas CP-violating contributions involving the $WW\gamma$ vertex remain non-zero. We propose that this is due to the fact that the neutral TGCs lack a C-even, P-odd component, as opposed to the charged TGCs which contain a C-even, P-odd form factor in addition to their C-odd, P-even component, as we will show in detail. Although this result does not establish a general selection rule, it illustrates the importance of distinguishing between different realisations of CP violation. A detailed investigation of the proposed EDM protection, including axial gauge-fermion couplings and fully off-shell TGC structures, will be presented in a forthcoming study.}

%The importance of this distinction is illustrated by the recent full three-loop calculation of Ref.~\cite{Banno:2026hsc}, in which heavy SU(2)$_L$ multiplets couple to the SM only through gauge interactions, as the inert scalars do here, but carry a CP-violating phase in the relative scalar and pseudoscalar parts of their Yukawa couplings, i.e.\ a C-even, P-odd source. The resulting electron EDM arises at three loops through the CP-odd $WW\gamma$ vertex, the electroweak Weinberg operator $\epsilon^{abc}W^a_{\mu\nu}W^{b\nu}{}_\rho\widetilde W^{c\rho\mu}$, while the contributions associated with the $ZZ\gamma$ and $\gamma\gamma\gamma$ vertices cancel between opposite orientations of the internal fermion loop. This pattern follows the discrete-symmetry structure of the vertices: the neutral TGCs are all C-odd and have no C-even, P-odd component, whereas the charged TGCs contain the C-even, P-odd structures $\widetilde\kappa_V$ and $\widetilde\lambda_V$ (Sec.~\ref{sec:cviolating}). A gauge-mediated dark sector therefore does feed EDMs when its CP violation is P-odd. A sector of scalars and gauge bosons alone cannot generate the $\epsilon$-tensor structure of the electroweak Weinberg operator, and this is the starting point of our conjecture. Ref.~\cite{Banno:2026hsc} contains no C-odd, P-even source, and so does not test the conjecture directly. A detailed investigation of the proposed EDM protection, including axial gauge-fermion couplings and fully off-shell TGC structures, will be presented in a forthcoming study.

Notwithstanding this, such CP violation can be transmitted to the SM sector through gauge interactions and contribute to baryogenesis, an example of such mechanism is shown in Ref.~\cite{Keus:2024khd}.
This should come as no surprise since Sakharov conditions explicitly require C and CP violation as essential requirements of baryogenesis (as opposed to P and CP violation). A detailed analysis of this CP violation transfer mechanism is under-way and will be presented in an upcoming publication.

It is important to note that when the extended inert sector contains a doublet and a singlet scalar~\cite{Azevedo:2018fmj}, instead of two doublet scalars, the amount of CP-violation is inevitably reduced due to the presence of the singlet, as established in Ref.~\cite{Keus:2019szx}.
This is a consequence of the presence of
fewer number of inert states contributing to the relevant processes, along with diluted $ZS_iS_j$ couplings (since the singlet has no direct couplings to the SM gauge bosons).
In the analysis that follows, we use the example of the aforementioned 3HDM which is the most minimal model contributing to many signatures we present. We will, however, mention where a model with smaller particle content can contribute.

The remainder of the paper is organised as follows.
In Sec.~\ref{sec:edmops} we review the EDM experiments and operators, establishing that all of them are C-even and P-odd. In Sec.~\ref{sec:cviolating} we show, at the level of general principles, that CP violation confined to a bosonic sector of scalars and gauge fields must instead be C-odd and P-even, identify the associated C-odd, P-even form factors in triple gauge boson processes, and formulate our conjecture on their embedding into the full electroweak theory. Sec.~\ref{sec:co-occurrence} discusses C-selection-rule tests: transitions of the same state into final states of mutually incompatible C eigenvalue. Sec.~\ref{sec:multipoles} places this result in the broader context of static multipole moments, identifies the class of moments that shares the discrete-symmetry quantum numbers of our CP violation, and explains why static moments of fermions cannot provide C-selection-rule tests.
Sec.~\ref{sec:model} presents the I(2+1)HDM as a concrete realisation of this class of CP violation, and Sec.~\ref{sec:conclusions} concludes.

%%%%%%%%%%%%%%%%%%%%%%%%%%%%%%%%%%%%
%%%%%%%%%%%%%%%%%%%%%%%%%%%%%%%%%%%%
\section{EDM operators: P violating, C conserving}
\label{sec:edmops}
%%%%%%%%%%%%%%%%%%%%%%%%%%%%%%%%%%%%
%%%%%%%%%%%%%%%%%%%%%%%%%%%%%%%%%%%%

Permanent EDMs are searched for through spin-precession measurements in external electric and magnetic fields. For a spin-$\nicefrac{1}{2}$ particle, an EDM induces the interaction $- d\:\!\vec E\cdot \frac{\vec S}{S}$,
which shifts the magnetic Larmor precession frequency $\omega_L$ according to $\hbar\omega_{\uparrow\downarrow}=2(\mu B\pm dE)$, so that the observed total spin-precession frequency is
\be
\omega_{\mathrm{obs}} = \omega_L + \delta \omega_{\mathrm{EDM}}, ~~~~~\mbox{with}~~~
\omega_L= \frac{2\mu B}{\hbar} ~~~\mbox{and} ~~~ \delta \omega_{\mathrm{EDM}}=\pm \frac{2dE}{\hbar},
\ee
where the sign depends on the relative orientation of $\vec E$ and $\vec B$ fields. Reversing the electric field isolates the EDM contribution while cancelling the ordinary magnetic precession.

EDM searches employ a range of complementary experimental systems. Neutron EDM, $d_n$, experiments use stored ultracold neutrons in static electric and magnetic fields, with Ramsey spectroscopy and co-magnetometry providing precise measurements of spin precession; the current limit is $|d_n|<1.8\times10^{-26}\:\!\mathrm{e \:\! cm}$ at $90\%$ C.L.~\cite{Abel:2020pzs}. The strongest constraints on the electron EDM, $d_e$, arise from paramagnetic polar molecules such as ThO and HfF$^+$, where very large internal effective electric fields strongly enhance sensitivity to $d_e$~\cite{ACME:2018yjb,Roussy:2022cmp}. The present bound is $|d_e|<4.1\times10^{-30}\:\!\mathrm{e \:\! cm}$ at $90\%$ C.L.~\cite{Roussy:2022cmp}, with future molecular-ion experiments expected to improve this sensitivity further~\cite{Hutzler:2020lmj}. Diamagnetic systems such as $^{199}$Hg, $^{129}$Xe and $^{225}$Ra instead probe predominantly hadronic and nuclear sources of CP violation, including nuclear Schiff moments and quark chromo-electric dipole moments~\cite{Engel:2013lsa}. Finally, EDMs of charged particles such as protons, deuterons and muons can be probed in storage rings, where an EDM induces an out-of-plane spin precession and hence a measurable vertical polarisation~\cite{CPEDM:2019nwp,Muong-2:2008ebm}.

Although neutron, atomic, molecular, and storage-ring experiments probe different combinations of effective operators, they all measure the same underlying observable: a flavour-diagonal EDM.
For a fermion, the corresponding relativistic interaction is
\be
\mathcal{L}_{\rm EDM}
=
-\frac{i}{2}d\:\!
\bar\psi\sigma^{\mu\nu}\gamma_5\psi
F_{\mu\nu},
\label{eq:edmlag}
\ee
which is C-even, P-odd, and T-odd. Consequently, a non-zero EDM signals P and T violation and, assuming CPT invariance, CP violation.
At the energies relevant for EDM searches, the underlying CP-violating interactions are mapped onto hadronic, nuclear, atomic, and molecular matrix elements through an effective field theory expansion~\cite{Engel:2013lsa,Pospelov:2005pr}.
The complete list of flavour-diagonal EDM operators can be found in Ref.~\cite{Engel:2013lsa}.\footnote{
We find it instructive to show the electron EDM calculation in detail in Appendix~\ref{app:EDMs} as they are the most relevant for the scalar extensions, and are experimentally most constraining.}
It is important to note that every operator in this list is C-even and P-odd.
Consequently, a source of CP violation that is C-odd and P-even cannot generate any of these operators on its own. In a theory in which C and P are separately conserved apart from such a source, this is an exact consequence of the symmetry structure, rather than a suppression by loops or small couplings. In the SM, C and P are additionally violated by the chiral gauge interactions of the fermions; whether this can convert a C-odd, P-even source into an EDM is the subject of Sec.~\ref{sec:2V*-cancellation}.

In the following sections we identify the processes through which this class of CP violation can instead be probed.

%%%%%%%%%%%%%%%%%%%%%%%%%%%%%%%%%%%%
%%%%%%%%%%%%%%%%%%%%%%%%%%%%%%%%%%%%
\section{Bosonic operators: C violating, P conserving}
\label{sec:cviolating}
%%%%%%%%%%%%%%%%%%%%%%%%%%%%%%%%%%%%
%%%%%%%%%%%%%%%%%%%%%%%%%%%%%%%%%%%%

In a renormalisable theory of scalar fields transforming under a gauge group $G$, together with the corresponding gauge fields, with all dimension-four-or-lower terms consistent with the gauge symmetry retained, the gauge-kinetic terms conserve C, P and T, while the scalar potential, through complex mass parameters or self-couplings, can introduce CP-violating terms.
Crucially, however, parity remains exactly conserved irrespective of the scalar potential: one can consistently assign $\mathrm{P}=+1$ to all scalar fields and $\mathrm{P}=-1$ to all gauge fields,\footnote{The assignment $\mathrm{P}=-1$ for gauge fields follows from the parity properties of the four-velocity $u^\mu$ in the current $j^\mu=\rho u^\mu$ that sources $\mathcal L=-j^\mu A_\mu$.} and verify that P is conserved at every vertex once the orbital angular momentum of each transition has been taken into account.\footnote{An illustrative example is a vertex with one gauge boson and two scalars, such as the standard $ZHA$ vertex in the CP-conserving 2HDM. One can verify that parity is conserved after taking into account the $(-1)^\ell$ factor, where $\ell=1$ is the orbital angular momentum involved in the transition from a spin-1 particle to a pair of spin-0 particles.}
Any CP violation generated by the scalar potential can therefore be interpreted as a violation of C rather than P~\cite{Haber:2022gsn}.\footnote{There exist also the topological term $\theta F\tilde F$ which is P-odd and CP-odd, and is not relevant to our discussion.}

To illustrate this, consider a generic neutral scalar $S$, which need not possess a definite CP quantum number, decaying into a pair of gauge bosons $VV$ through a loop of charged scalars. The resulting effective interaction contains the P-conserving operator $SF_{\mu\nu}F^{\mu\nu}$, whereas the P-violating operator $SF_{\mu\nu}\widetilde F^{\mu\nu}$ cannot be generated by purely scalar and gauge interactions at any order in perturbation theory. This follows from the absence of parity-odd tensor structures, in particular the Levi-Civita tensor $\epsilon^{\mu\nu\alpha\beta}$, in the underlying interactions. Consequently, even in the presence of CP violation in the scalar sector, the resulting bosonic amplitudes conserve P, implying that any CP violation must originate from C violation. The same reasoning applies to the scalar-loop contributions to the triple-gauge vertices discussed in the following sections.

%%%%%%%%%%%%%%%%%%%%%%%%%%%%%%%%%%%%
\subsection{Triple gauge boson processes}
%%%%%%%%%%%%%%%%%%%%%%%%%%%%%%%%%%%%

The triple gauge boson processes were first introduced in~\cite{Gaemers:1978hg,Renard:1981es,Hagiwara:1986vm} and have been studied extensively ever since, for early studies see~\cite{Baur:1992cd,Hagiwara:1994pw,Wudka:1994ny,Aihara:1995iq,CDF:1994bpj,Ellison:1998uy,Gounaris:1999kf,Gounaris:2000dn}.

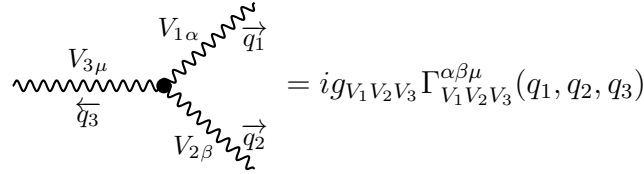
\begin{figure}[h!]
\begin{center}
\begin{tikzpicture}[thick,scale=1.0]
\draw[photon] (0,0) -- node[black,above,xshift=0cm,yshift=0cm] {$V_{3\:\!\mu}$} (2,0);
\draw (0,0)  node[black,above,xshift=1cm,yshift=-0.7cm] {$\overleftarrow{{q_3}}$} (2,0);
\draw[photon] (2,0) -- node[black,above,xshift=-0.4cm,yshift=-0.1cm] {$V_{1\:\!\alpha}$} (3.2,1.1);
\draw[photon] (2,0) -- node[black,above,xshift=-0.2cm,yshift=-0.6cm] {$V_{2\:\!\beta}$} (3.2,-1.1);
\draw (3.2,1.1)  node[black,above,xshift=-0.0cm,yshift=-0.8cm] {$\overrightarrow{{q_1}}$} (4,1.1);
\draw  (3.2,-1.1)  node[black,above,xshift=0.0cm,yshift=0.1cm] {$\overrightarrow{{q_2}}$} (4,-1.1);
\fill[black] (2,0) circle (0.1cm);
\draw (0,0)  node[black,above,xshift=6.0cm,yshift=-0.4cm] {\normalsize $= ig_{V_1V_2V_3} \:\!\Gamma^{\alpha\beta\mu}_{V_1V_2V_3}({q_1},{q_2},q_3)$} (2,0);
\end{tikzpicture}\\[-2mm]
\caption{\small The general triple gauge boson vertex $V_1V_2V_3$.}
\label{fig:V1V2V3}
\end{center}
\end{figure}

For a triple-gauge-boson interaction $V_1V_2V_3$, with associated momenta $q_1,q_2,q_3$, the most general Lorentz-invariant vertex, represented schematically in Fig.\ref{fig:V1V2V3}, can be decomposed into a basis of independent tensor structures multiplied by the corresponding form factors.
This parametrisation is particularly suitable for the analysis here, where the new-physics scale is set by the masses of the additional scalar states and need not be parametrically larger than the characteristic momentum scales of the process.
In this regime, a truncated effective-field-theory expansion in local higher-dimensional operators may not provide an accurate description of the full momentum dependence of the vertex~\cite{Buchmuller:1985jz,Grzadkowski:2010es}. The form-factor approach, by contrast, allows this momentum dependence to be retained explicitly. We therefore formulate our analysis directly in terms of the triple-gauge-boson form factors rather than basing it on a truncated set of gauge-invariant effective operators. For completeness, the corresponding effective-operator description for fully off-shell vector bosons is presented in Appendix~\ref{app:fully-off-shell}.

%%%%%%%%%%%%%%%%%%%%%%%%%%%%%%%%%%%%
\subsubsection{The $WWV$ vertex with $V=Z$ or $\gamma$}
%%%%%%%%%%%%%%%%%%%%%%%%%%%%%%%%%%%%

The effective Lagrangian generating the triple gauge vertices involving two $W$ bosons is
\bea
\nicefrac{1}{g_{WWV}} \mathcal L_{WWV} &=&
ig_1^V\big(W^\dagger_{\rho\sigma}W^\rho V^\sigma - W^\dagger_\rho V_\sigma W^{\rho\sigma}\big)
+ i\kappa_V\:\!W^\dagger_\rho W_\sigma V^{\rho\sigma}
+ \nicefrac{i}{m_W^2}\lambda_VW^\dagger_{\lambda\rho}W^\rho_{\ \sigma}V^{\sigma\lambda}
\nonumber\\[1mm]
&& - g_4^V\:\!W^\dagger_\rho W_\sigma(\partial^\rho V^\sigma+\partial^\sigma V^\rho)
+ g_5^V\:\!\epsilon^{\rho\sigma\kappa\lambda} \big(W^\dagger_\rho \partial_\kappa W_\sigma - \partial_\kappa W^\dagger_\rho  W_\sigma  \big) V_\lambda
\nonumber\\[1mm]
&&
+ i\widetilde\kappa_V\:\!W^\dagger_\rho W_\sigma\widetilde V^{\rho\sigma}
+ \nicefrac{i}{m_W^2}\widetilde\lambda_VW^\dagger_{\lambda\rho}W^\rho_{\ \sigma}\widetilde V^{\sigma\lambda},
\label{eq:L-WWV}
\eea
where
$ V_{\mu\nu} = \partial_\mu V_\nu - \partial_\nu V_\mu $ (and similarly for $ W_{\mu \nu} $),
$ \widetilde V_{\mu \nu} = {1\over2} \epsilon_{\mu\nu\alpha\beta}
V^{\alpha \beta }$, $g_{WW \gamma}=-e$, $g_{WWZ}=-e \cot \theta_W $, and $\theta_W$ denotes the weak mixing angle.
In the SM at tree level the values of these couplings are
$\kappa_V = g_1^V = 1$, and
$\lambda_V = \widetilde \lambda_V = \widetilde \kappa_V = g_4^V = g_5^V = 0$.
The couplings $g_4^V$, $\widetilde \kappa_V$, and $\widetilde \lambda_V$ are CP violating.
All other couplings are CP conserving.\footnote{One can establish the C, P, T properties of the form factors by applying the C, P, T  transformations to the effective Lagrangian.}

\begin{figure}[h!]
\begin{center}
\begin{tikzpicture}[thick,scale=1.0]
\draw[photon] (0,0) -- node[black,above,xshift=0cm,yshift=0cm] {$Z^*_\mu$} (2,0);
\draw (0,0)  node[black,above,xshift=1cm,yshift=-0.8cm] {$\overleftarrow{{q_3}}$} (2,0);
\draw[dashed] (2,0) -- node[black,above,xshift=-0.2cm,yshift=-0.1cm] {$S_i$} (3.2,1.1);
\draw[dashed] (3.2,1.1) -- node[black,above,xshift=0.3cm,yshift=-0.3cm] {$S^\pm_k$} (3.2,-1.1);
\draw[dashed] (2,0) -- node[black,above,xshift=-0.2cm,yshift=-0.5cm] {$S_j$} (3.2,-1.1);
\draw[photon] (3.2,1.1) -- node[black,above,xshift=-0.1cm,yshift=0cm] {$W_\alpha$} (5,1.1);
\draw (3.2,1.1)  node[black,above,xshift=1.4cm,yshift=-0.8cm] {$\overrightarrow{{q_1}}$} (5,1.1);
\draw[photon] (3.2,-1.1) -- node[black,above,xshift=-0.1cm,yshift=-0.7cm] {$W_\beta$} (5,-1.1);
\draw  (3.2,-1.1)  node[black,above,xshift=1.4cm,yshift=0.05cm] {$\overrightarrow{{q_2}}$} (5,-1.1);
\draw (0,0)  node[black,above,xshift=1.5cm,yshift=-2cm] {$\mathrm{\circled{a}}$} (2,0);
\end{tikzpicture}
\hspace{10mm}
\begin{tikzpicture}[thick,scale=1.0]
\draw[photon] (0,0) -- node[black,above,xshift=0cm,yshift=0cm] {$Z^*_\mu$} (2,0);
\draw (0,0)  node[black,above,xshift=1cm,yshift=-0.8cm] {$\overleftarrow{{q_3}}$} (2,0);
\draw[dashed] (2,0) -- node[black,above,xshift=-0.2cm,yshift=-0.1cm] {$S^\pm_i$} (3.2,1.1);
\draw[dashed] (3.2,1.1) -- node[black,above,xshift=0.3cm,yshift=-0.3cm] {$S_k$} (3.2,-1.1);
\draw[dashed] (2,0) -- node[black,above,xshift=-0.2cm,yshift=-0.7cm] {$S^\pm_j$} (3.2,-1.1);
\draw[photon] (3.2,1.1) -- node[black,above,xshift=-0.1cm,yshift=0cm] {$W_\alpha$} (5,1.1);
\draw (3.2,1.1)  node[black,above,xshift=1.4cm,yshift=-0.8cm] {$\overrightarrow{{q_1}}$} (5,1.1);
\draw[photon] (3.2,-1.1) -- node[black,above,xshift=-0.1cm,yshift=-0.7cm] {$W_\beta$} (5,-1.1);
\draw  (3.2,-1.1)  node[black,above,xshift=1.4cm,yshift=0.05cm] {$\overrightarrow{{q_2}}$} (5,-1.1);
\draw (0,0)  node[black,above,xshift=1.5cm,yshift=-2cm] {$\mathrm{\circled{b}}$} (2,0);
\end{tikzpicture}
\\[3mm]
\begin{tikzpicture}[thick,scale=1.0]
\draw[photon] (0,0) -- node[black,above,xshift=0cm,yshift=0cm] {$\gamma^*_\mu$} (2,0);
\draw (0,0)  node[black,above,xshift=1cm,yshift=-0.8cm] {$\overleftarrow{{q_3}}$} (2,0);
\draw[dashed] (2,0) -- node[black,above,xshift=-0.2cm,yshift=-0.1cm] {$S^\pm_i$} (3.2,1.1);
\draw[dashed] (3.2,1.1) -- node[black,above,xshift=0.3cm,yshift=-0.4cm] {$S_j$} (3.2,0.4);
\draw[photon] (3.2,0.4) node[black,above,xshift=-0.01cm,yshift=-0.7cm] {$W^\pm$} arc (90:270:0.4cm) ;
\draw[dashed] (3.2,0.4) node[black,above,xshift=0.7cm,yshift=-0.8cm] {$S^\pm_k$} arc (90:-90:0.4cm) ;
\draw[dashed] (3.2,-0.4) -- node[black,above,xshift=0.3cm,yshift=-0.3cm] {$S_\ell$} (3.2,-1.1);
\draw[dashed] (2,0) -- node[black,above,xshift=-0.2cm,yshift=-0.7cm] {$S^\pm_i$} (3.2,-1.1);
\draw[photon] (3.2,1.1) -- node[black,above,xshift=-0.1cm,yshift=0cm] {$W_\alpha$} (5,1.1);
\draw (3.2,1.1)  node[black,above,xshift=1.4cm,yshift=-0.8cm] {$\overrightarrow{{q_1}}$} (5,1.1);
\draw[photon] (3.2,-1.1) -- node[black,above,xshift=-0.1cm,yshift=-0.7cm] {$W_\beta$} (5,-1.1);
\draw  (3.2,-1.1)  node[black,above,xshift=1.4cm,yshift=0.05cm] {$\overrightarrow{{q_2}}$} (5,-1.1);
\draw (0,0)  node[black,above,xshift=1.5cm,yshift=-2cm] {$\mathrm{\circled{c}}$} (2,0);
\end{tikzpicture}
\hspace{10mm}
\begin{tikzpicture}[thick,scale=1.0]
\draw[photon] (0,0) -- node[black,above,xshift=0cm,yshift=0cm] {$\gamma^*_\mu$} (2,0);
\draw (0,0)  node[black,above,xshift=1cm,yshift=-0.8cm] {$\overleftarrow{{q_3}}$} (2,0);
\draw[dashed] (2,0) -- node[black,above,xshift=-0.2cm,yshift=-0.1cm] {$S^\pm_i$} (3.2,1.1);
\draw[dashed] (3.2,1.1) -- node[black,above,xshift=0.3cm,yshift=-0.4cm] {$S_j$} (3.2,0.4);
\draw[photon] (3.2,0.4) node[black,above,xshift=-0.1cm,yshift=-0.7cm] {$Z$} arc (90:270:0.4cm) ;
\draw[dashed] (3.2,0.4) node[black,above,xshift=0.7cm,yshift=-0.8cm] {$S_k$} arc (90:-90:0.4cm) ;
\draw[dashed] (3.2,-0.4) -- node[black,above,xshift=0.3cm,yshift=-0.3cm] {$S_\ell$} (3.2,-1.1);
\draw[dashed] (2,0) -- node[black,above,xshift=-0.2cm,yshift=-0.7cm] {$S^\pm_i$} (3.2,-1.1);
\draw[photon] (3.2,1.1) -- node[black,above,xshift=-0.1cm,yshift=0cm] {$W_\alpha$} (5,1.1);
\draw (3.2,1.1)  node[black,above,xshift=1.4cm,yshift=-0.8cm] {$\overrightarrow{{q_1}}$} (5,1.1);
\draw[photon] (3.2,-1.1) -- node[black,above,xshift=-0.1cm,yshift=-0.7cm] {$W_\beta$} (5,-1.1);
\draw  (3.2,-1.1)  node[black,above,xshift=1.4cm,yshift=0.05cm] {$\overrightarrow{{q_2}}$} (5,-1.1);
\draw (0,0)  node[black,above,xshift=1.5cm,yshift=-2cm] {$\mathrm{\circled{d}}$} (2,0);
\end{tikzpicture}\\[-1mm]
\caption{\small The diagrams for the $WWV$ vertices ($V=\gamma, Z$ and $i\neq j\neq k \neq \ell$) with the effective vertex coefficient $i g_{WWV}\Gamma^{\alpha\beta\mu}_V({q_1},{q_2},q_3)$. Note that in the absence of the extra internal loop in $WW\gamma$ diagrams, the opposite CP phase of the $W^{\pm} S_i^\mp S_j$ vertices would cancel each other and demote the process into a CP conserving one.}
\label{fig:WWV}
\end{center}
\end{figure}
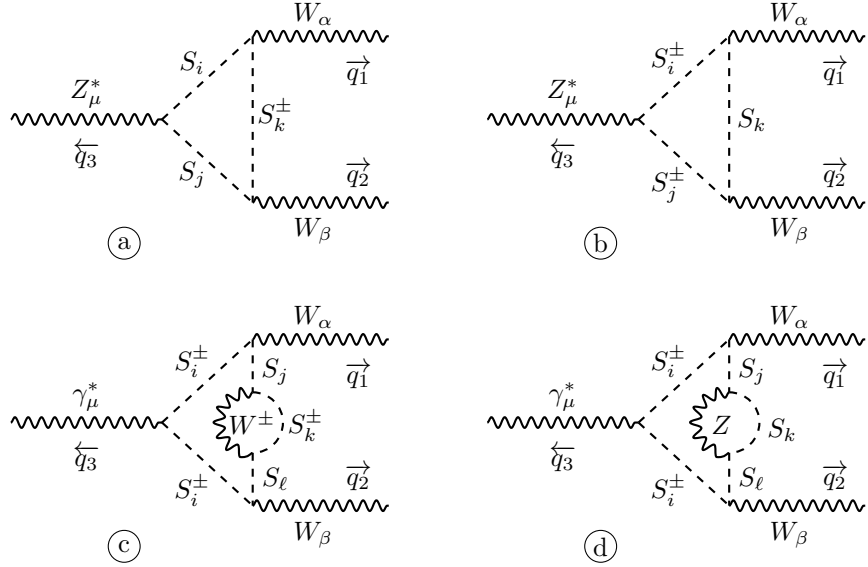

In momentum space, the corresponding effective vertex can be expressed as
\bea
\Gamma^{\alpha\beta\mu}_{WWV}({q_1},{q_2},q_3) &= &
f_1^{WWV}({q_1}- {q_2})^\mu g^{\alpha\beta}
- \nicefrac{1}{m_W^2}f_2^{WWV}({q_1}- {q_2})^\mu q_3^\alpha q_3^\beta
- f_3^{WWV}\!\left(q_3^\alpha g^{\mu\beta}-q_3^\beta g^{\mu\alpha}\right) \notag\\[1mm]
&&- i f_4^{WWV}\!\left(q_3^\alpha g^{\mu\beta}+q_3^\beta g^{\mu\alpha}\right)
+ i f_5^{WWV}\:\! \epsilon^{\mu\alpha\beta\rho}({q_1}- {q_2})_\rho
+ f_6^{WWV}\:\! \epsilon^{\mu\alpha\beta\rho}q_{3\:\!\rho}
\notag\\[1mm]
&& + \nicefrac{1}{m_W^2}f_7^{WWV}({q_1}-{q_2})^\mu \:\! \epsilon^{\alpha\beta\rho\sigma}q_{3\:\!\rho}({q_1}-{q_2})_\sigma \:\!,
\label{eq:Gamma-WWV}
\eea
where ${q_1}, \:\!{q_2},\:\!{q_3}$ represent the outgoing momenta, as shown in Fig.\ref{fig:WWV}.

One can calculate the contribution of the lowest-dimension operators in Eq.\eqref{eq:L-WWV} to the seven form factors which are\footnote{Contributions from higher dimensional operators provide the $q_3^2$ dependence of the form factors.}
\bea
&&
(\text{P-even, C-even})
\quad
f_1^{WWV} = g_1^V + \nicefrac{q_3^2}{2 m_W^2}\lambda_V \:\!,
\quad
f_2^{WWV}  = \lambda_V  \:\!,
\quad
f_3^{WWV}  = g_1^V + \kappa_V +  \lambda_V \:\!,
\nonumber\\
&&
(\text{P-even, C-odd})
\quad ~
f_4^{WWV}  = g_4^V \:\!,
\nonumber\\[2mm]
&&
(\text{P-odd, C-odd})
\quad ~~
f_5^{WWV} =  g_5^V \:\!,
\label{eq:fs-WWV}\\[2mm]
&&
(\text{P-odd, C-even})
\quad ~
f_6^{WWV}  = \widetilde \kappa_V -\widetilde  \lambda_V \:\!,
\quad
f_7^{WWV}  = -\nicefrac{1}{2}\widetilde \lambda_V \:\!.
\nonumber
\eea
For $V=\gamma$, these coefficients have relation to physical quantities (static moments) as follows:
\be
\label{eq:WWV-moments}
\begin{tabular}{ll}
$q_W = e g_1^\gamma$
&
(electric monopole, \textit{i.e.}, charge)
\\[3mm]
$\mu_W = { e \over 2 m_W } ( g_1^\gamma + \kappa_\gamma + \lambda_\gamma)$
&
(magnetic dipole)
\\[3mm]
$d_W = { e \over 2 m_W } ( \tilde \kappa_\gamma + \tilde \lambda_\gamma )$
&
(electric dipole)
\\[3mm]
$Q^m_W = - { e \over m_W^2 } ( \tilde \kappa_\gamma - \tilde \lambda_\gamma ) $
&
(magnetic quadrupole)
\\[3mm]
$Q^e_W = - { e \over m_W^2 } ( \kappa_\gamma - \lambda_\gamma )$
&
(electric quadrupole)
\\
\end{tabular}
\ee
%where the first term with $g_1^\gamma=1$ determines the charge of the $W$ boson $q_W$ is the so-called \textit{minimal}coupling term.
%and the second coefficient $ \kappa_\gamma$ is conventionally called the 'anomalous' magnetic moment of the $W$. This term and the third coefficient $\lambda_\gamma$ are related to the magnetic moment $\mu_W$ and the electric quadrupole moment $Q^e_W$ of the $W$

Fig.\ref{fig:WWV} shows the diagrams contributing to the C-violating P-conserving $WWZ$ and $WW\gamma$ processes.
The $WWZ$ process receives its lowest-order C-violating P-conserving contributions at one loop, requiring at least two non-identical neutral and one charged scalar as shown in diagram~{\footnotesize \circled{a}}, and at least two non-identical charged and one neutral scalar as shown in diagram~{\footnotesize \circled{b}}.
Therefore, beyond the particle content of the SM, a CP-violating $\Z_2$-symmetric dark sector is needed with at least one scalar doublet and one complex scalar singlet for producing diagram~{\footnotesize \circled{a}}, and at least two extra scalar doublets for diagram~{\footnotesize \circled{b}}.

The C-violating P-conserving $WW\gamma$ process first appears at two loops. This is due to
the unavoidable diagonal couplings of the photon leg, leading to cancellation of the CP phase of the $W^{+} S_i^- S_j$ and $W^{-} S_i^+ S_j$ vertices, and therefore demoting the process into a CP conserving one.
The extra internal loop in the $WW\gamma$ diagrams avoids such cancellation, and
requires at least two non-identical neutral and two non-identical charged scalars as shown in diagram~{\footnotesize \circled{c}}, and at least three non-identical neutral and one charged scalar as shown in diagram~{\footnotesize \circled{d}}.
Therefore, beyond the particle content of the SM, the $WW\gamma$ process requires a CP-violating $\Z_2$-symmetric dark sector with at least two extra scalar doublets, \textit{i.e.}, our 3HDM framework discussed in Sec.~\ref{sec:model} for generating diagram~{\footnotesize \circled{c}} and one doublet plus one singlet for diagram~{\footnotesize \circled{d}}.

%%%%%%%%%%%%%%%%%%%%%%%%%%%%%%%%%%%%
\subsubsection{The $ZZV$ vertex with $V=Z$ or $\gamma$}
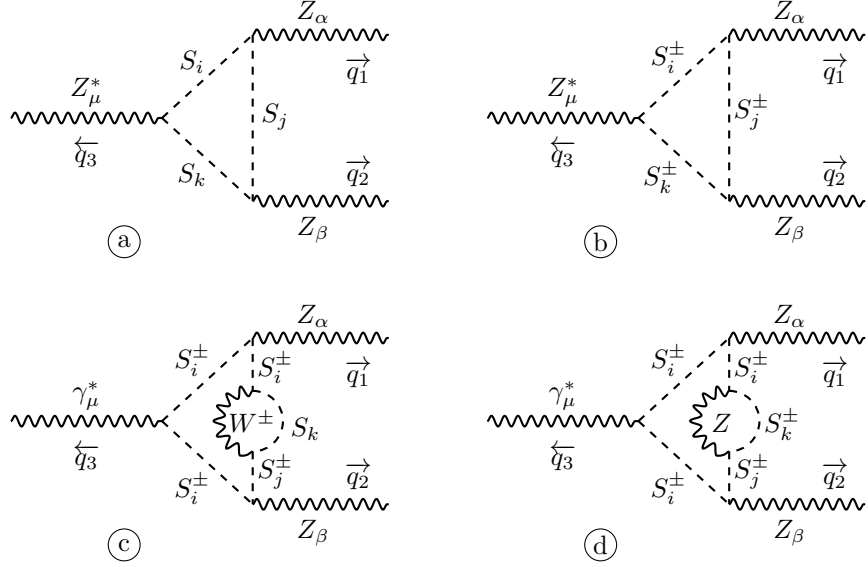
\begin{figure}[h!]
\begin{center}
\begin{tikzpicture}[thick,scale=1.0]
\draw[photon] (0,0) -- node[black,above,xshift=0cm,yshift=0cm] {$Z^*_\mu$} (2,0);
\draw (0,0)  node[black,above,xshift=1cm,yshift=-0.8cm] {$\overleftarrow{{q_3}}$} (2,0);
\draw[dashed] (2,0) -- node[black,above,xshift=-0.2cm,yshift=-0.1cm] {$S_i$} (3.2,1.1);
\draw[dashed] (3.2,1.1) -- node[black,above,xshift=0.3cm,yshift=-0.3cm] {$S_j$} (3.2,-1.1);
\draw[dashed] (2,0) -- node[black,above,xshift=-0.2cm,yshift=-0.5cm] {$S_k$} (3.2,-1.1);
\draw[photon] (3.2,1.1) -- node[black,above,xshift=-0.1cm,yshift=0cm] {$Z_\alpha$} (5,1.1);
\draw (3.2,1.1)  node[black,above,xshift=1.4cm,yshift=-0.8cm] {$\overrightarrow{{q_1}}$} (5,1.1);
\draw[photon] (3.2,-1.1) -- node[black,above,xshift=-0.1cm,yshift=-0.7cm] {$Z_\beta$} (5,-1.1);
\draw  (3.2,-1.1)  node[black,above,xshift=1.4cm,yshift=0.05cm] {$\overrightarrow{{q_2}}$} (5,-1.1);
\draw (0,0)  node[black,above,xshift=1.5cm,yshift=-2cm] {$\mathrm{\circled{a}}$} (2,0);
\end{tikzpicture}
\hspace{10mm}
\begin{tikzpicture}[thick,scale=1.0]
\draw[photon] (0,0) -- node[black,above,xshift=0cm,yshift=0cm] {$Z^*_\mu$} (2,0);
\draw (0,0)  node[black,above,xshift=1cm,yshift=-0.8cm] {$\overleftarrow{{q_3}}$} (2,0);
\draw[dashed] (2,0) -- node[black,above,xshift=-0.2cm,yshift=-0.1cm] {$S^\pm_i$} (3.2,1.1);
\draw[dashed] (3.2,1.1) -- node[black,above,xshift=0.3cm,yshift=-0.3cm] {$S^\pm_j$} (3.2,-1.1);
\draw[dashed] (2,0) -- node[black,above,xshift=-0.3cm,yshift=-0.6cm] {$S^\pm_k$} (3.2,-1.1);
\draw[photon] (3.2,1.1) -- node[black,above,xshift=-0.1cm,yshift=0cm] {$Z_\alpha$} (5,1.1);
\draw (3.2,1.1)  node[black,above,xshift=1.4cm,yshift=-0.8cm] {$\overrightarrow{{q_1}}$} (5,1.1);
\draw[photon] (3.2,-1.1) -- node[black,above,xshift=-0.1cm,yshift=-0.7cm] {$Z_\beta$} (5,-1.1);
\draw  (3.2,-1.1)  node[black,above,xshift=1.4cm,yshift=0.05cm] {$\overrightarrow{{q_2}}$} (5,-1.1);
\draw (0,0)  node[black,above,xshift=1.5cm,yshift=-2cm] {$\mathrm{\circled{b}}$} (2,0);
\end{tikzpicture}
\\[3mm]
\begin{tikzpicture}[thick,scale=1.0]
\draw[photon] (0,0) -- node[black,above,xshift=0cm,yshift=0cm] {$\gamma^*_\mu$} (2,0);
\draw (0,0)  node[black,above,xshift=1cm,yshift=-0.8cm] {$\overleftarrow{{q_3}}$} (2,0);
\draw[dashed] (2,0) -- node[black,above,xshift=-0.2cm,yshift=-0.1cm] {$S^\pm_i$} (3.2,1.1);
\draw[dashed] (3.2,1.1) -- node[black,above,xshift=0.3cm,yshift=-0.4cm] {$S^\pm_i$} (3.2,0.4);
\draw[photon] (3.2,0.4) node[black,above,xshift=-0.01cm,yshift=-0.7cm] {$W^\pm$} arc (90:270:0.4cm) ;
\draw[dashed] (3.2,0.4) node[black,above,xshift=0.7cm,yshift=-0.8cm] {$S_k$} arc (90:-90:0.4cm) ;
\draw[dashed] (3.2,-0.4) -- node[black,above,xshift=0.3cm,yshift=-0.3cm] {$S^\pm_j$} (3.2,-1.1);
\draw[dashed] (2,0) -- node[black,above,xshift=-0.2cm,yshift=-0.7cm] {$S^\pm_i$} (3.2,-1.1);
\draw[photon] (3.2,1.1) -- node[black,above,xshift=-0.1cm,yshift=0cm] {$Z_\alpha$} (5,1.1);
\draw (3.2,1.1)  node[black,above,xshift=1.4cm,yshift=-0.8cm] {$\overrightarrow{{q_1}}$} (5,1.1);
\draw[photon] (3.2,-1.1) -- node[black,above,xshift=-0.1cm,yshift=-0.7cm] {$Z_\beta$} (5,-1.1);
\draw  (3.2,-1.1)  node[black,above,xshift=1.4cm,yshift=0.05cm] {$\overrightarrow{{q_2}}$} (5,-1.1);
\draw (0,0)  node[black,above,xshift=1.5cm,yshift=-2cm] {$\mathrm{\circled{c}}$} (2,0);
\end{tikzpicture}
\hspace{10mm}
\begin{tikzpicture}[thick,scale=1.0]
\draw[photon] (0,0) -- node[black,above,xshift=0cm,yshift=0cm] {$\gamma^*_\mu$} (2,0);
\draw (0,0)  node[black,above,xshift=1cm,yshift=-0.8cm] {$\overleftarrow{{q_3}}$} (2,0);
\draw[dashed] (2,0) -- node[black,above,xshift=-0.2cm,yshift=-0.1cm] {$S^\pm_i$} (3.2,1.1);
\draw[dashed] (3.2,1.1) -- node[black,above,xshift=0.3cm,yshift=-0.4cm] {$S^\pm_i$} (3.2,0.4);
\draw[photon] (3.2,0.4) node[black,above,xshift=-0.1cm,yshift=-0.7cm] {$Z$} arc (90:270:0.4cm) ;
\draw[dashed] (3.2,0.4) node[black,above,xshift=0.7cm,yshift=-0.8cm] {$S^\pm_k$} arc (90:-90:0.4cm) ;
\draw[dashed] (3.2,-0.4) -- node[black,above,xshift=0.3cm,yshift=-0.3cm] {$S^\pm_j$} (3.2,-1.1);
\draw[dashed] (2,0) -- node[black,above,xshift=-0.2cm,yshift=-0.7cm] {$S^\pm_i$} (3.2,-1.1);
\draw[photon] (3.2,1.1) -- node[black,above,xshift=-0.1cm,yshift=0cm] {$Z_\alpha$} (5,1.1);
\draw (3.2,1.1)  node[black,above,xshift=1.4cm,yshift=-0.8cm] {$\overrightarrow{{q_1}}$} (5,1.1);
\draw[photon] (3.2,-1.1) -- node[black,above,xshift=-0.1cm,yshift=-0.7cm] {$Z_\beta$} (5,-1.1);
\draw  (3.2,-1.1)  node[black,above,xshift=1.4cm,yshift=0.05cm] {$\overrightarrow{{q_2}}$} (5,-1.1);
\draw (0,0)  node[black,above,xshift=1.5cm,yshift=-2cm] {$\mathrm{\circled{d}}$} (2,0);
\end{tikzpicture}\\[-1mm]
\caption{\small The diagrams for the $ZZV$ vertices ($V=\gamma, Z$ and $i\neq j\neq k$) with the effective vertex coefficient $i g_{ZZV}\Gamma^{\alpha\beta\mu}_V({q_1},{q_2},q_3)$. Note that in the absence of the extra internal loop in $ZZ\gamma$ diagrams, the opposite CP phase of the $Z S_i^\pm S^\mp_j$ vertices would cancel each other and demote the process into a CP conserving one.}
\label{fig:ZZV}
\end{center}
\end{figure}

%%%%%%%%%%%%%%%%%%%%%%%%%%%%%%%%%%%%
As a consequence of Bose statistics, the $Z$ and $\gamma$ self-couplings vanish identically, when all three particles are on-shell.
The general form of the vertex function with a pair of on-shell $ZZ$ and one off-shell boson $V=Z,\gamma$ which is coupled to a conserved current can be constructed.
The effective Lagrangian generating such vertices is
\bea
\L_{ZZV} &=& \nicefrac{e}{\mzd} \biggl[
-f_4^{ZZ\gamma} (\partial_\mu F^{\mu \beta}) Z_\alpha
( \partial^\alpha Z_\beta)
-f_4^{ZZZ} (\partial_\mu Z^{\mu \beta})  Z_\alpha
( \partial^\alpha Z_\beta) \nonumber\\
&& ~~~ ~~~~
+f_5^{ZZ\gamma} (\partial^\sigma F_{\sigma \mu}) \wtil{Z}^{\mu \beta} Z_\beta
+f_5^{ZZZ} (\partial^\sigma Z_{\sigma \mu})  \wtil{Z}^{\mu \beta} Z_\beta \biggr]\:\!,
\label{eq:L-ZZV}
\eea
where $\wtil{Z}_{\mu \nu}=\nicefrac{1}{2} \epsilon_{\mu \nu \rho \sigma}Z^{\rho
\sigma}$ with
$Z_{\mu\nu}=\partial_\mu Z_\nu -\partial_\nu Z_\mu$ and similarly for
the photon tensor $F_{\mu\nu}$.
The most general form of the $ZZV^*$ vertex function can be written as
\be
\Gamma^{\alpha \beta \mu}_{ZZ V} ({q_1}, {q_2}, q_3)
= \nicefrac{i (q_3^2-m_V^2)}{\mzd}
\biggl[ - f_4^{ZZV}  (q_3^\alpha g^{\mu \beta}+q_3^\beta g^{\mu \alpha})
-f_5^{ZZV}  \epsilon^{\mu \alpha \beta \rho}({q_1} - {q_2})_\rho \biggr]\:\!,
\label{eq:Gamma-ZZV}
\ee
where ${q_1}, \:\!{q_2},\:\! q_3$ represent the outgoing momenta, as shown in Fig.\ref{fig:ZZV}.
Note that the coupling
$f_4^{ZZV}$ is C-odd, P-even while $f_5^{ZZV}$ is C-odd, P-odd and hence CP-conserving.

Fig.\ref{fig:ZZV} shows the diagrams contributing to the C-violating P-conserving $ZZZ$ and $ZZ\gamma$ processes.
These processes are absent at tree-level in the SM.
The $ZZZ$ process receives its lowest-order C-violating P-conserving contributions at one loop, requiring at least three non-identical neutral scalars in diagram~{\footnotesize \circled{a}}, and at least three non-identical charged scalars in diagram~{\footnotesize \circled{b}}.
Therefore, beyond the particle content of the SM, a CP-violating $\Z_2$-symmetric dark sector is needed with at least two extra scalar doublets to generate diagram~{\footnotesize \circled{a}}, and at least three extra scalar doublets for diagram~{\footnotesize \circled{b}}.\footnote{Note, however, that allowing for an extra bosonic loop in the propagator of one of the charged scalars in diagram~{\footnotesize \circled{b}}, reduces the minimum requirement of three non-identical charged scalars by one, hence, the minimal model producing this diagram becomes our 3HDM.}

The C-violating P-conserving $ZZ\gamma$ process first appears at two loops. This is due to the fact that in the absence of the extra internal loop the CP phase of the $Z S_i^+ S^-_j$ and $Z S_i^- S^+_j$ vertices would cancel each other and demote the process into a CP conserving one. To avoid such cancellation,
one requires at least two non-identical charged and one neutral scalar to generate diagram~{\footnotesize \circled{c}}, and at least three non-identical charged scalars for diagram~{\footnotesize \circled{d}}.
Therefore, beyond the particle content of the SM, the $ZZ\gamma$ process requires a CP-violating $\Z_2$-symmetric dark sector with at least two extra scalar doublets in diagram~{\footnotesize \circled{c}}, and at least three extra scalar doublets in diagram~{\footnotesize \circled{d}}.

%%%%%%%%%%%%%%%%%%%%%%%%%%%%%%%%%%%%
\subsubsection{The $Z\gamma V$ vertex with $V=Z$ or $\gamma$}
%%%%%%%%%%%%%%%%%%%%%%%%%%%%%%%%%%%%

\begin{figure}[h!]
\begin{center}
\begin{tikzpicture}[thick,scale=1.0]
\draw[photon] (0,0) -- node[black,above,xshift=0cm,yshift=0cm] {$Z^*_\mu$} (2,0);
\draw (0,0)  node[black,above,xshift=1cm,yshift=-0.8cm] {$\overleftarrow{{q_3}}$} (2,0);
\draw[dashed] (2,0) -- node[black,above,xshift=-0.2cm,yshift=-0.1cm] {$S^\pm_i$} (3.2,1.1);
\draw[dashed] (3.2,1.1) -- node[black,above,xshift=0.3cm,yshift=-0.4cm] {$S^\pm_i$} (3.2,0.4);
\draw[photon] (3.2,0.4) node[black,above,xshift=-0.01cm,yshift=-0.7cm] {$W^\pm$} arc (90:270:0.4cm) ;
\draw[dashed] (3.2,0.4) node[black,above,xshift=0.7cm,yshift=-0.8cm] {$S_k$} arc (90:-90:0.4cm) ;
\draw[dashed] (3.2,-0.4) -- node[black,above,xshift=0.3cm,yshift=-0.3cm] {$S^\pm_j$} (3.2,-1.1);
\draw[dashed] (2,0) -- node[black,above,xshift=-0.2cm,yshift=-0.7cm] {$S^\pm_j$} (3.2,-1.1);
\draw[photon] (3.2,1.1) -- node[black,above,xshift=-0.1cm,yshift=0cm] {$Z_\alpha$} (5,1.1);
\draw (3.2,1.1)  node[black,above,xshift=1.4cm,yshift=-0.8cm] {$\overrightarrow{{q_1}}$} (5,1.1);
\draw[photon] (3.2,-1.1) -- node[black,above,xshift=-0.1cm,yshift=-0.7cm] {$\gamma_\beta$} (5,-1.1);
\draw  (3.2,-1.1)  node[black,above,xshift=1.4cm,yshift=0.05cm] {$\overrightarrow{{q_2}}$} (5,-1.1);
\draw (0,0)  node[black,above,xshift=1.5cm,yshift=-2cm] {$\mathrm{\circled{a}}$} (2,0);
\end{tikzpicture}
\hspace{10mm}
\begin{tikzpicture}[thick,scale=1.0]
\draw[photon] (0,0) -- node[black,above,xshift=0cm,yshift=0cm] {$\gamma^*_\mu$} (2,0);
\draw (0,0)  node[black,above,xshift=1cm,yshift=-0.8cm] {$\overleftarrow{{q_3}}$} (2,0);
\draw[dashed] (2,0) -- node[black,above,xshift=-0.2cm,yshift=-0.1cm] {$S^\pm_i$} (3.2,1.1);
\draw[dashed] (3.2,1.1) -- node[black,above,xshift=0.3cm,yshift=-0.4cm] {$S^\pm_j$} (3.2,0.4);
\draw[photon] (3.2,0.4) node[black,above,xshift=-0.01cm,yshift=-0.7cm] {$W^\pm$} arc (90:270:0.4cm) ;
\draw[dashed] (3.2,0.4) node[black,above,xshift=0.7cm,yshift=-0.8cm] {$S_k$} arc (90:-90:0.4cm) ;
\draw[dashed] (3.2,-0.4) -- node[black,above,xshift=0.3cm,yshift=-0.3cm] {$S^\pm_i$} (3.2,-1.1);
\draw[dashed] (2,0) -- node[black,above,xshift=-0.2cm,yshift=-0.7cm] {$S^\pm_i$} (3.2,-1.1);
\draw[photon] (3.2,1.1) -- node[black,above,xshift=-0.1cm,yshift=0cm] {$Z_\alpha$} (5,1.1);
\draw (3.2,1.1)  node[black,above,xshift=1.4cm,yshift=-0.8cm] {$\overrightarrow{{q_1}}$} (5,1.1);
\draw[photon] (3.2,-1.1) -- node[black,above,xshift=-0.1cm,yshift=-0.7cm] {$\gamma_\beta$} (5,-1.1);
\draw  (3.2,-1.1)  node[black,above,xshift=1.4cm,yshift=0.05cm] {$\overrightarrow{{q_2}}$} (5,-1.1);
\draw (0,0)  node[black,above,xshift=1.5cm,yshift=-2cm] {$\mathrm{\circled{b}}$} (2,0);
\end{tikzpicture}\\[-1mm]
\caption{\small The diagrams for the $Z \gamma V$ vertices ($V=\gamma, Z$ and $i\neq j\neq k $) with the effective vertex coefficient $i g_{Z \gamma V}\Gamma^{\alpha\beta\mu}_V({q_1},{q_2},q_3)$. Note that both these processes require at least two loops for a non-zero C-violating P-conserving CP violation, and at least two extra Higgs doublets beyond the particle content of the SM.}
\label{fig:ZgammaV}
\end{center}
\end{figure}
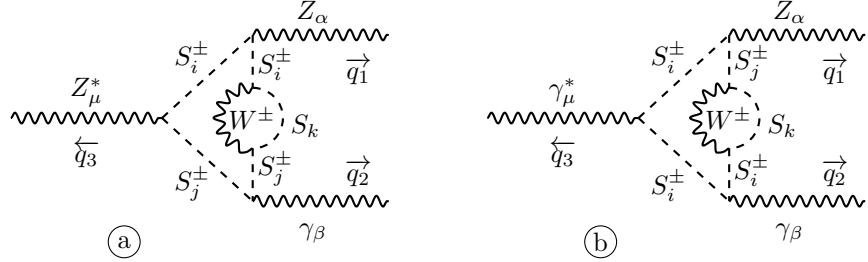

The effective Lagrangian generating the $Z\gamma V$ vertices ($V=\gamma, Z$) is
\bea
\L_{Z \gamma V}
&=& \nicefrac{e}{\mzd} \biggl[
- h_1^{Z\gamma\gamma} \left(\partial^\sigma F_{\sigma \mu}\right) Z_\beta F^{\mu \beta}
- h_1^{Z\gamma Z} (\partial^\sigma Z_{\sigma \mu}) Z_\beta F^{\mu \beta}
\nonumber\\
&&
~~~~~~~
- \nicefrac{1}{\mzd}h_2^{Z\gamma\gamma} \left(\partial_\alpha \partial_\beta
\partial^\rho F_{\rho \mu} \right) Z^\alpha F^{\mu \beta}
-\nicefrac{1}{\mzd}h_2^{Z\gamma Z} \left(\partial_\alpha \partial_\beta
(\square +\mzd) Z_\mu\right)  Z^\alpha F^{\mu \beta}
\nonumber \\[2mm]
&&
~~~~~~~
-  h_3^{Z\gamma\gamma} \left(\partial_\sigma F^{\sigma \rho}\right) Z^\alpha
 \wtil{F}_{\rho \alpha}
- h_3^{Z\gamma Z}  \left(\partial_\sigma Z^{\sigma \rho}\right) Z^\alpha
 \wtil{F}_{\rho \alpha}
\nonumber \\[2mm]
&&
~~~~~~~
+\nicefrac{1}{2\mzd}h_4^{Z\gamma\gamma}\left(\square \partial^\sigma
F^{\rho \alpha}\right) Z_\sigma \wtil{F}_{\rho \alpha }
+\nicefrac{1}{2 \mzd}h_4^{Z\gamma Z} \left((\square +\mzd) \partial^\sigma
Z^{\rho \alpha}\right)  Z_\sigma \wtil{F}_{\rho \alpha }
\biggr] \:\!.
\label{eq:L-ZgammaV}
\eea
Similar to the previous case, the most general form of the $Z \gamma V$ vertex function with off-shell $V=Z,\gamma$ which is coupled to a conserved current is
\bea
\Gamma^{\alpha \beta \mu}_{Z\gamma V} ({q_1}, {q_2}, q_3)
&=& \nicefrac{ i (q_3^2-m_V^2)}{\mzd}
\biggl[ h_1^{Z\gamma V} ({q^\mu_2} g^{\alpha \beta}-{q^\alpha_2} g^{\mu \beta} )
+ \nicefrac{1}{\mzd}h_2^{Z\gamma V} q_3^\alpha [ (q_3{q_2}) g^{\mu \beta}- {q^\mu_2} q_3^\beta ]
\nonumber \\
&& ~~~~~~~~~~~~~~~~~
- h_3^{Z\gamma V} \epsilon^{\mu \alpha \beta \rho} {q_{2\:\!\rho}}
~-~\nicefrac{1}{\mzd}h_4^{Z\gamma V} q_3^\alpha \epsilon^{\mu \beta \rho
\sigma}q_{3\:\!\rho} {q_{2\:\!\sigma}} \biggr] \:\! , \label{eq:Gamma-ZgammaV}
\eea
where ${q_1},\:\!{q_2},\:\! q_3$ represent the outgoing momenta, as also shown in Fig.\ref{fig:ZgammaV}.
Note again that the couplings
$h_1^V,h_2^V$ are C-odd P-even, and $h_3^V,h_4^V$ are C-odd P-odd and hence CP-even.
For $V=Z$ these coefficients refer to the following \textit{transition moments} where $ E_\gamma $ is
the photon energy:
\be
\begin{tabular}{ll}
$q_Z = 0$
&
(electric monopole, \textit{i.e.}, charge)
\\[2mm]
$\mu_Z = { - e \over \sqrt{2} \:\! m_Z} { E_\gamma^2 \over m_Z^2} \left( h_1^Z - h_2^Z \right)$
&
(magnetic dipole)
\\[3mm]
$d_Z = { - e \over \sqrt{2} \:\! m_Z} { E_\gamma^2 \over m_Z^2} \left( h_3^Z - h_4^Z \right)$
&
(electric dipole)
\\[3mm]
$Q^m_Z = { 2 \sqrt{10} \:\! e \over m_Z^2} h_3^Z $
&
(magnetic quadrupole)
\\[3mm]
$Q^e_Z = { 2 \sqrt{10} \:\! e \over m_Z^2} h_1^Z $
&
(electric quadrupole)
\\
\end{tabular}
\ee

Fig.\ref{fig:ZgammaV} shows the diagrams contributing to the C-violating P-conserving $Z\gamma Z$ and $Z\gamma\gamma$ processes.
These processes are absent at tree-level in the SM, and receive their lowest-order C-violating P-conserving contributions at two loop, requiring at least two non-identical charged and one neutral scalar to generate diagrams~{\footnotesize \circled{a}} and~{\footnotesize \circled{b}}.
Therefore, beyond the particle content of the SM, a CP-violating $\Z_2$-symmetric dark sector is needed with at least two extra scalar doublets, as in our 3HDM framework.

%%%%%%%%%%%%%%%%%%%%%%%%%%%%%%%%%%%%
%%%%%%%%%%%%%%%%%%%%%%%%%%%%%%%%%%%%
\subsection{Embedding the TGC processes into EDM diagrams}
%Alternative title: No C-violation CP-violation when two gauge bosons are off-shell
\label{sec:2V*-cancellation}
%%%%%%%%%%%%%%%%%%%%%%%%%%%%%%%%%%%%
%%%%%%%%%%%%%%%%%%%%%%%%%%%%%%%%%%%%

The TGC form-factor parametrisations employed in this work are defined for the specific on-shell and off-shell configurations indicated for each vertex. In particular, the off-shell gauge-boson leg is assumed to couple to a conserved current, satisfying $q_\mu J^\mu=0$, so that tensor structures proportional to its momentum do not contribute to the amplitude. While this condition follows from electromagnetic current conservation for an off-shell photon, it represents an additional restriction for off-shell $Z$ and $W$ bosons.

Within these kinematic conditions, embedding the C-violating, P-conserving TGC structures into amplitudes involving only C- and P-conserving interactions, such as vectorial gauge--fermion couplings, cannot generate the C-even, P-odd structure required for an EDM. However, the SM also contains axial gauge--fermion interactions, which violate both C and P. Their inclusion renders an EDM contribution compatible with the overall discrete-symmetry quantum numbers, although it does not establish that such a contribution is non-zero.

Importantly, the reduced TGC form factors considered here do not provide a complete description of these higher-order embeddings. In particular, fully off-shell $V_1^\ast V_2^\ast V_3^\ast$ vertices may contain additional longitudinal and momentum-dependent structures that are absent or do not contribute under the restricted kinematic conditions adopted here~\cite{Gounaris:2000tb,Gavela:2014vra}. Whether these structures contribute to an EDM depends on their coefficients in the underlying theory and on the constraints imposed by gauge invariance, Bose symmetry and the complete fermionic amplitude.\footnote{Similarly, attaching additional gauge bosons to the triple-gauge subdiagram introduces higher-point interactions, including quartic gauge-boson amplitudes, which are not described by the reduced TGC form factors and must be considered separately.}

The same question has been studied for fermionic interactions that are T-odd and P-even, the class to which our CP violation belongs. Refs.~\cite{Conti:1992xn,Ramsey-Musolf:1999cub,Kurylov:2000ub} showed that such interactions do induce EDMs once they are dressed with the parity-violating weak interactions of the SM, and that the resulting EDM bounds are strong. Those results concern fermionic contact operators, whereas the sources considered here are purely bosonic and enter the SM sector only through gauge-boson vertices; they nevertheless show that the separation between C-odd, P-even sources and EDMs is not guaranteed by symmetry alone once the full electroweak theory is included. A sharper version of the question can be formulated in the SMEFT. At dimension six, all CP-odd bosonic operators, $Q_{\widetilde W}$, $Q_{\widetilde G}$, $Q_{\varphi\widetilde W}$, $Q_{\varphi\widetilde B}$, $Q_{\varphi\widetilde G}$ and $Q_{\varphi\widetilde WB}$, contain the Levi-Civita tensor and are P-odd. They therefore cannot be generated by loops of inert scalars and gauge bosons alone, and first arise at the lowest order at which a chiral SM fermion loop supplies the $\epsilon$-tensor. The EDM then requires at least one further loop.

We therefore put forward the following conjecture:
\begin{tcolorbox}
The C-odd, P-even CP violation originating in the inert scalar sector remains inaccessible to EDM observables even after the corresponding interactions are embedded into the complete electroweak theory.
\end{tcolorbox}
Establishing this conjecture requires demonstrating that potentially relevant contributions involving axial fermionic currents either vanish or cancel in the complete gauge-invariant amplitude, in contrast with the fermionic T-odd, P-even interactions of Refs.~\cite{Conti:1992xn,Ramsey-Musolf:1999cub,Kurylov:2000ub}. Should it fail, the arguments above imply that any EDM contribution arises only beyond the order at which the P-odd bosonic operators are first generated, and carries additional electroweak suppression. Such a demonstration may be obtained through an explicit calculation of the EDM form factor or through a more general analysis of the symmetry and tensor properties of the fully off-shell amplitudes. A systematic investigation of these contributions is in progress and will be presented in subsequent work.

%%%%%%%%%%%%%%%%%%%%%%%%%%%%%%%%%%%%
%%%%%%%%%%%%%%%%%%%%%%%%%%%%%%%%%%%%
\section{C-selection rules from co-occurrence}
\label{sec:co-occurrence}
%%%%%%%%%%%%%%%%%%%%%%%%%%%%%%%%%%%%
%%%%%%%%%%%%%%%%%%%%%%%%%%%%%%%%%%%%

So far, we have established that a non-zero $f^{V}_4$, $h^{V}_1$ or $h^{V}_2$ signals CP violation in a P-even, C-odd structure, without requiring simultaneous observation of any other channel. These form factors can also be induced at higher order by a P-odd CP-violating source combined with the parity-violating weak interactions of the SM; such contributions are, however, bounded by the EDM constraints on that source, so that a value significantly above this bound points to a genuinely P-even origin.
These form factors provide quantitative probes of C-violating, P-conserving interactions.
In this section, we identify signatures that require simultaneous observation of certain processes to establish the presence of C violation.

%%%%%%%%%%%%%%%%%%%%%%%%%%%%%%%%%%%%
%%%%%%%%%%%%%%%%%%%%%%%%%%%%%%%%%%%%
\subsection{Co-occurrence, not a single forbidden channel}
\label{sec:signatures2}
%%%%%%%%%%%%%%%%%%%%%%%%%%%%%%%%%%%%
%%%%%%%%%%%%%%%%%%%%%%%%%%%%%%%%%%%%

As discussed in Ref.~\cite{Haber:2022gsn}, simultaneous observation of suitably chosen processes can provide evidence for the violation of discrete symmetries without requiring an independent assignment of the corresponding quantum numbers to the initial state. In the present context, a particularly transparent example is the observation of the same neutral scalar mass eigenstate, $h$, in both of the following channels\footnote{Although the one-loop $h\to\gamma\gamma\gamma$ amplitude vanishes for purely vectorial fermion-loop contributions by Furry's theorem, the decay is not generally forbidden at higher orders in the electroweak theory. In the I(2+1)HDM, the C-odd, P-even contribution originating from the CP-violating inert scalar sector first arises at three-loop order. The three-photon decay is therefore unlikely to provide a competitive discovery channel, despite its particularly transparent interpretation as a C-selection-rule test.}
\begin{tcolorbox}
\be
h \to\gamma\gamma
\qquad\quad ~\text{and}\qquad\quad ~
h \to\gamma\gamma\gamma
\ee
\end{tcolorbox}
Since the two- and three-photon final states have opposite C eigenvalues, $\mathrm{C}(\gamma\gamma)=+1$ and $\mathrm{C}(\gamma\gamma\gamma)=-1$, these transitions cannot both conserve C for a single initial state. Their co-occurrence therefore constitutes a selection-rule test of charge-conjugation invariance, independently of any prior C assignment to $h$.

This observation should, however, be distinguished from the identification of a new source of C-violating CP violation. Since the weak interactions of the SM already violate C, the observation of both channels would not, by itself, establish that the violation originates in the extended scalar sector. Such an interpretation requires identifying contributions beyond the SM expectation and establishing their discrete-symmetry properties.

The same selection-rule argument can be extended to interactions involving neutral electroweak gauge bosons. In a C-conserving theory, a neutral scalar with a definite C eigenvalue cannot simultaneously couple to two- and three-vector final states with opposite C assignments. Candidate comparisons therefore include $h\to ZZ$, $h\to Z\gamma$ and $h\to\gamma\gamma$, together with the corresponding three-vector processes $h\to ZZZ$, $h\to ZZ\gamma$, $h\to Z\gamma\gamma$ and $h\to\gamma\gamma\gamma$. For the SM-like Higgs boson, the $Z$ bosons in the three-vector channels are necessarily off shell where required by kinematics. These comparisons illustrate the general C-selection-rule principle; establishing the observability of the dark-sector CP-violating contributions to these processes requires dedicated calculations of the corresponding amplitudes and SM backgrounds.

This strategy has well-established precedents. The decay $\pi^0\to3\gamma$ provides a classic test of C invariance, since the neutral pion has $\mathrm C=+1$, whereas the three-photon final state has $\mathrm C=-1$~\cite{ParticleDataGroup:2024cfk}. Similarly, para-positronium ($\mathrm C=+1$) decays predominantly into an even number of photons, while ortho-positronium ($\mathrm C=-1$) decays predominantly into an odd number, and searches for C-forbidden positronium decays continue to test this selection rule~\cite{Jasinska:2016qsf}. The proposed observables extend the same C-selection-rule principle to the scalar sector.

%%%%%%%%%%%%%%%%%%%%%%%%%%%%%%%%%%%%
%%%%%%%%%%%%%%%%%%%%%%%%%%%%%%%%%%%%
\section{Static multipole moments and the electron EDM}
\label{sec:multipoles}
%%%%%%%%%%%%%%%%%%%%%%%%%%%%%%%%%%%%
%%%%%%%%%%%%%%%%%%%%%%%%%%%%%%%%%%%%

Since the $f_4^V$, $h_1^V$ and $h_2^V$ type vertices and the co-occurrence processes discussed so far are loop-level probes, it is natural to ask whether simpler, single-particle observables, namely static multipole moments, could serve the same diagnostic purpose. In this section we classify the static moments by their discrete-symmetry quantum numbers, identify the class that shares those of our CP violation, and explain why static moments cannot provide the C-selection-rule tests of Sec.~\ref{sec:co-occurrence}.

Any static charge and current distribution can be expanded in electric and magnetic $2^\ell$-pole moments ($\ell=0,1,2,\ldots$). A complete description of the source, including its contact interactions with external currents, also requires the toroidal (anapole) family. As classical quantities, all three families are C-odd, since they are linear in the charge density or the current.
Electric moments have $(\mathrm{P},\mathrm{T})=((-1)^\ell,+)$, magnetic moments $(\mathrm{P},\mathrm{T})=((-1)^{\ell+1},-)$, and toroidal moments $(\mathrm{P},\mathrm{T})=((-1)^\ell,-)$.
By the Wigner-Eckart theorem, a diagonal static moment of a particle with spin $\vec J$ is proportional to a rank-$\ell$ tensor built from $\vec J$, which has $\mathrm{P}=+1$ and $\mathrm{T}=(-1)^\ell$, and exists only for $\ell\le 2J$. Whenever the classical (P,T) properties of a moment differ from those of the spin tensor, the corresponding interaction operator violates that symmetry. The C quantum number of the operator is then inferred assuming CPT-invariance.\footnote{One must distinguish between the transformation properties of the multipole moments, those of the corresponding interaction operators, and the C eigenvalues of physical states; these are not interchangeable. In particular, all classical moments are C-odd, while the interaction operators of the electric and magnetic families are C-even and those of the toroidal family are C-odd.} The resulting general rule is given in Tab.~\ref{tab:multipole-rule}, and the explicit assignments up to $\ell=3$ are listed in Tab.~\ref{tab:multipole-list}.

\begin{table}[h!]
\centering
\begin{tabular}{|l|c|c||c|c|l|}\hline
Family & Moment (P,T) & C & P & T & Couples to \Tstrut\Bstrut\\
\hline
Electric $2^\ell$ & $((-1)^\ell,\,+)$ & $+$ & $(-1)^\ell$ & $(-1)^\ell$ & $\nabla^{\ell-1}\vec E$ \Tstrut\\[1.5mm]
Magnetic $2^\ell$ & $((-1)^{\ell+1},\,-)$ & $+$ & $(-1)^{\ell+1}$ & $(-1)^{\ell+1}$ & $\nabla^{\ell-1}\vec B$ \\[1.5mm]
Toroidal $2^\ell$ & $((-1)^\ell,\,-)$ & $-$ & $(-1)^\ell$ & $(-1)^{\ell+1}$ & $\nabla^{\ell-1}\vec j_{\rm ext}$ (contact) \Bstrut\\[1mm]
\hline
\end{tabular}\\[-1mm]
\caption{\small C, P and T assignments of the interaction operators of the electric, magnetic and toroidal $2^\ell$-pole moments of a particle with spin $J$, together with the (P,T) of the corresponding classical moment. The (P,T) of the operator follows from comparing the classical moment with the rank-$\ell$ spin tensor (P$=+1$, T$=(-1)^\ell$), and the C of the operator then follows assuming CPT-invariance.}
\label{tab:multipole-rule}
\end{table}

\begin{table}[h!]
\centering
\begin{tabular}{|c|l|c|c|c|c|}\hline
$\ell$ & Moment & C & P & T & $J_{\rm min}$ \Tstrut\Bstrut\\
\hline
0 & Charge (charge radius) & $+$ & $+$ & $+$ & 0 \Tstrut\\[1mm]
0 & Magnetic monopole & $+$ & $-$ & $-$ & 0 \\[1mm]
1 & Electric dipole $d$ (Schiff moment) & $+$ & $-$ & $-$ & $1/2$ \\[1mm]
1 & Magnetic dipole $\mu$ & $+$ & $+$ & $+$ & $1/2$ \\[1mm]
1 & Anapole $a$ (toroidal dipole) & $-$ & $-$ & $+$ & $1/2$ \\[1mm]
2 & Electric quadrupole $Q_E$ & $+$ & $+$ & $+$ & 1 \\[1mm]
2 & Magnetic quadrupole $Q_M$ & $+$ & $-$ & $-$ & 1 \\[1mm]
2 & Toroidal quadrupole $T_2$ & $-$ & $+$ & $-$ & 1 \\[1mm]
3 & Electric octupole & $+$ & $-$ & $-$ & $3/2$ \\[1mm]
3 & Magnetic octupole & $+$ & $+$ & $+$ & $3/2$ \\[1mm]
3 & Toroidal octupole & $-$ & $-$ & $+$ & $3/2$ \Bstrut\\
\hline
\end{tabular}\\[-1mm]
\caption{\small C, P and T assignments of the interaction operators of the static electromagnetic multipole moments up to $\ell=3$. $J_{\rm min}$ is the smallest spin supporting the moment, from $\ell\le 2J$; a spin-0 particle therefore supports only the rank-0 moments, \textit{i.e.} the electric charge (and in a P- and T-violating theory, a magnetic charge) independently of any other discrete symmetry.}
\label{tab:multipole-list}
\end{table}

The pattern alternates with $\ell$ and produces four classes:
\begin{itemize}
\item \textit{ordinary} moments, \textit{i.e.,} even-$\ell$ electric and odd-$\ell$ magnetic (charge, magnetic dipole, electric quadrupole, \ldots), with $\mathrm{C}=\mathrm{P}=\mathrm{T}=+1$, present in any theory and carrying no information about new physics,
\item \textit{EDM-type} moments, \textit{i.e.,} odd-$\ell$ electric and even-$\ell$ magnetic (electric dipole, magnetic quadrupole, electric octupole, \ldots), with $\mathrm{C}=+1$ and $\mathrm{P}=\mathrm{T}=-1$, which vanish unless the dynamics violates P and T,
\item \textit{anapole-type} moments, \textit{i.e.,} odd-$\ell$ toroidal (anapole, toroidal octupole, \ldots), with $\mathrm{C}=\mathrm{P}=-1$ and $\mathrm{T}=+1$, which are CP-even and require only the P (and C) violation already present in the weak interactions,
\item \textit{toroidal moments with even-$\ell$} (toroidal quadrupole, \ldots), with $\mathrm{C}=\mathrm{T}=-1$ and $\mathrm{P}=+1$, which are CP-odd while conserving P.
\end{itemize}

The EDM belongs to the C-even, P-odd and T-odd class. The CP violation originating in the inert scalar sector is instead C-odd and P-even, \textit{i.e.}, it shares the discrete-symmetry quantum numbers of the last class above, which first appear at rank two and hence require a particle with $J\ge1$. In nuclear physics, interactions of this class are known as time-reversal-violating, parity-conserving (TVPC) interactions; their EDM constraints were discussed in Sec.~\ref{sec:2V*-cancellation}. For the $W$ boson this class is represented by the $g_4^\gamma$ coupling of Eq.(\ref{eq:fs-WWV}), which is C-odd and P-even, just as the anapole-type and EDM-type classes are represented in Eq.(\ref{eq:WWV-moments}) by $g_5^\gamma$ and by $\tilde\kappa_\gamma$, $\tilde\lambda_\gamma$, respectively.\footnote{Being toroidal, these structures are contact interactions with the external current: for an on-shell photon, electromagnetic gauge invariance requires $g_4^\gamma$ and $g_5^\gamma$ to vanish~\cite{Hagiwara:1986vm}.
This is why they do not appear among the static moments of Eq.(\ref{eq:WWV-moments}). The identification of $g_4^\gamma$ with the toroidal quadrupole should be understood as a correspondence of discrete-symmetry class rather than a one-to-one dictionary.} In the I(2+1)HDM, this coupling first arises at two loops, as discussed in Sec.~\ref{sec:cviolating}. The distinction between the EDM-type and the even-$\ell$ toroidal classes underlies the EDM-protection mechanism proposed in this work. Whether it extends to the complete electroweak theory, including parity-violating fermionic interactions, remains the subject of our conjecture.

For a spin-zero particle, the Wigner--Eckart theorem forbids permanent static electromagnetic multipole moments of rank $\ell\ge1$, independently of the discrete symmetries of the underlying theory. In particular, the inert scalars of the I(2+1)HDM cannot possess permanent electric dipole moments, nor any other static electromagnetic moment of rank $\ell\ge1$. This does not exclude electromagnetic transition form factors or higher-point interactions, through which CP violation in the scalar sector may manifest itself; the triple-gauge structures of Sec.~\ref{sec:cviolating} provide examples of such observables.

%%%%%%%%%%%%%%%%%%%%%%%%%%%%%%%%%%%%%%%%%%%%%%%%%%%%%%
\subsection{Why static moments do not provide C-selection-rule tests}
\label{subsec:single-particle}
%%%%%%%%%%%%%%%%%%%%%%%%%%%%%%%%%%%%%%%%%%%%%%%%%%%%%%

The simultaneous observation of different electromagnetic multipole moments of a single particle does not provide the same type of C-selection-rule test as the co-occurrence of transitions into final states with incompatible C eigenvalues. Such a test requires an initial state of definite C and a transition into final states whose C eigenvalues can be compared with it. Static moments of fermions fail on both counts. Charged fermions are not C eigenstates, whereas neutral self-conjugate (Majorana) fermions are, but their diagonal moments are restricted by the Majorana condition itself. Moreover, a static moment is a property of a single state rather than a transition. We discuss the two cases in turn, using the relation between the bilinears of a fermion field and of its charge conjugate $\psi^c=C\bar\psi^{\,T}$, which after normal ordering reads
\bea
\bar\psi^c\,\Gamma\,\psi^c=\eta_\Gamma\,\bar\psi\,\Gamma\,\psi\,,\qquad
\eta_\Gamma=\begin{cases}+1\,, & \Gamma=1,\ i\gamma_5,\ \gamma^\mu\gamma_5\,,\\ -1\,, & \Gamma=\gamma^\mu,\ \sigma^{\mu\nu},\ \sigma^{\mu\nu}\gamma_5\,.\end{cases}
\label{eq:etaGamma}
\eea
For a spin-$\tfrac12$ particle, the condition $\ell\le 2J$ restricts the diagonal moments to the charge, magnetic dipole, electric dipole and anapole, \textit{i.e.} to the four elastic form factors $F_1$, $F_2$, $F_3$ and $F_A$, multiplying $\gamma^\mu$, $\sigma^{\mu\nu}q_\nu$, $\sigma^{\mu\nu}\gamma_5 q_\nu$ and $(q^2\gamma^\mu-\slashed{q}\,q^\mu)\gamma_5$, respectively~\cite{Nowakowski:2004cv,Giunti:2014ixa}.

\textit{Charged fermions.} An electron is not a C eigenstate, since charge conjugation maps it onto a positron, whose moments are fixed by those of the electron through CPT.\footnote{Rewriting the Lagrangian in terms of $\psi^c$ with Eq.~(\ref{eq:etaGamma}) shows that, relative to its own spin, the antiparticle carries $q_{\bar\psi}=-q_\psi$, $\mu_{\bar\psi}=-\mu_\psi$, $d_{\bar\psi}=-d_\psi$ and $a_{\bar\psi}=+a_\psi$; in general, its moment is that of the particle multiplied by $-\mathrm{C}$, with C the operator quantum number of Tab.~\ref{tab:multipole-rule}. These relations are guaranteed by CPT and hold whether or not C or CP is conserved; in particular $d_{\bar\psi}=-d_\psi$ holds although the EDM operator is CP-odd. Comparisons of $g_{e^+}$ with $g_{e^-}$, or of $\mu_{\bar p}$ with $\mu_p$, are therefore tests of CPT rather than of C.} The existence of its magnetic dipole moment therefore imposes no C assignment that could be contradicted by the observation of an EDM. Moreover, the electron EDM is extracted from the electromagnetic form factor of a charged fermion rather than from a transition between states of definite C: an isolated on-shell electron cannot radiate a real photon while remaining on shell, and EDM measurements instead probe the effective dipole interaction through processes involving external electromagnetic fields. The C-selection-rule argument based on incompatible final-state eigenvalues is therefore not applicable to the electron EDM.

\textit{Neutral fermions.} For a self-conjugate Majorana fermion, $\chi=\chi^c$, Eq.~(\ref{eq:etaGamma}) gives $\bar\chi\,\Gamma\,\chi=-\bar\chi\,\Gamma\,\chi=0$ for every structure with $\eta_\Gamma=-1$. The vector and tensor bilinears $\bar\chi\gamma^\mu\chi$, $\bar\chi\sigma^{\mu\nu}\chi$ and $\bar\chi\sigma^{\mu\nu}\gamma_5\chi$ therefore vanish identically, and so do the diagonal charge, magnetic dipole and electric dipole moments~\cite{Kayser:1982br,Kayser:1984ge}. This is an algebraic consequence of the Majorana condition on the field, and holds independently of whether the underlying interactions conserve C, P or CP. The axial-vector bilinear has $\eta_\Gamma=+1$ and survives, so the anapole is the only diagonal electromagnetic form factor of a Majorana fermion. Off-diagonal bilinears $\bar\chi_i\,\Gamma\,\chi_j$ with $i\neq j$ are not subject to this constraint, so transition magnetic and electric dipole moments between different Majorana mass eigenstates may exist.

Thus, neither the simultaneous observation of different static moments of a charged fermion nor the pattern of diagonal moments of a Majorana fermion provides the co-occurrence test considered here. The distinguishing feature of the proposed selection-rule tests is instead the observation of transitions of the same initial state into final states with incompatible C assignments, as shown in the two- and three-photon decay channels of Sec.~\ref{sec:co-occurrence}.

%%%%%%%%%%%%%%%%%%%%%%%%%%%%%%%%%%%%
%%%%%%%%%%%%%%%%%%%%%%%%%%%%%%%%%%%%
\section{A concrete realisation of dark CP violation}
\label{sec:model}
%%%%%%%%%%%%%%%%%%%%%%%%%%%%%%%%%%%%
%%%%%%%%%%%%%%%%%%%%%%%%%%%%%%%%%%%%

We now present a minimal phenomenological model realising the exact bosonic C-violating P-conserving sources.
%At least three Higgs doublets are required: a $\mathbb Z_2$-symmetric potential built from just two doublets reduces to the ordinary 2HDM, in which the absence of a protecting symmetry beyond the active/inert split allows CP violation to reach the fermion sector, as noted before.
A $\mathbb Z_2$-symmetric 3HDM potential can be written as $V_{\rm 3HDM}=V_0+V_{\mathbb Z_2}$, with
\bea
V_0 &= &- \displaystyle \sum_{i=1}^3 \mu_i^2(\phi_i^\dagger\phi_i)
+ \sum_{i,j=1}^{3} \lambda_{ij} (\phi_i^\dagger\phi_i)(\phi_j^\dagger\phi_j)
+ \sum_{\substack{i,j=1 \\ i\ne j}}^{3}  \lambda'_{ij}(\phi_i^\dagger\phi_j)(\phi_j^\dagger\phi_i)
\:\! ,  \nonumber\\
V_{\mathbb Z_2} &= &-\mu_{12}^2(\phi_1^\dagger\phi_2) + \lambda_1(\phi_1^\dagger\phi_2)^2+\lambda_2(\phi_2^\dagger\phi_3)^2+\lambda_3(\phi_3^\dagger\phi_1)^2 + \cdots + \text{h.c.}\:\! , \label{eq:vz2}
\eea
where $V_0$ is invariant under any phase rotation and $V_{\Z_2}$ is a collection of
extra terms which realise the $\Z_2$ symmetry group, ensuring the potential respects this symmetry and no larger one~\cite{Ivanov:2011ae,Ivanov:2012hc}.
The three doublets $\phi_{1,2,3}$ carry $\mathbb Z_2$ charges $(-,-,+)$, respectively, and can be expressed as
\be
\phi_1= \doublet{$\begin{scriptsize}$ H^+_1 $\end{scriptsize}$}{\frac{H_1+iA_1}{\sqrt{2}}},\quad
\phi_2= \doublet{$\begin{scriptsize}$ H^+_2 $\end{scriptsize}$}{\frac{H_2+iA_2}{\sqrt{2}}}, \quad
\phi_3= \doublet{$\begin{scriptsize}$ G^+ $\end{scriptsize}$}{\frac{v+h+iG^0}{\sqrt{2}}},
\label{eq:doublets}
\ee
with the $\mathbb Z_2$-odd doublets as \emph{inert} ($\langle\phi_1\rangle=\langle\phi_2\rangle=0$) and the $\mathbb Z_2$-even doublet as \emph{active} ($\langle\phi_3\rangle=\nicefrac{v}{\sqrt2}$) playing the role of the SM Higgs doublet, where $h$ is the 125 GeV SM-like Higgs and $G^\pm,G^0$ the Goldstone bosons. Because the vacuum respects the $\mathbb Z_2$ symmetry, the active and inert sectors do not mix: $h$ retains exactly the tree-level couplings of the SM Higgs boson, CP violation is confined to the inert sector where it appears explicitly through the complex parameters of $V_{\Z_2}$.

Diagonalising the resulting charged and neutral mass matrices yields four CP-mixed neutral eigenstates $S_{1,2,3,4}$ and two charged mass eigenstates $S_1^\pm,S_2^\pm$,
\be
\left( \begin{array} {c}
S_1\\ S_2 \\ S_3 \\S_4 \end{array} \right) = \mathcal{R}^n
\left( \begin{array} {c}
H_1 \\ A_1 \\ H_2  \\A_2 \end{array} \right) ,
\qquad
\left( \begin{array} {c}
S^\pm_1\\ S^\pm_2 \end{array} \right) = \mathcal{R}^c
\left( \begin{array} {c}
H^\pm_1\\ H^\pm_2  \end{array} \right) ,
\ee
where the rotation matrices $\mathcal{R}^n$ and $\mathcal{R}^c$ are derived from the potential.
For the discussion here, the relevant terms in the Lagrangian are
\bea
\mathcal{L}_{\rm gauge} \; &\supset & \; g_{ZS_iS_j} Z_\mu (S_i \partial^\mu S_j - S_j \partial^\mu S_i)
+  g_{ZS^\pm_iS^\mp_j} Z_\mu (S^\pm_i \partial^\mu S^\mp_j - S^\pm_j \partial^\mu S^\mp_i)  \nonumber\\
&& +
g_{W^\pm S^\mp_i S_j} W^\pm_\mu (S^\mp_i \partial^\mu S_j - S_j \partial^\mu S^\mp_i)\,.
\eea

This spectrum and coupling structure is what populates the bosonic loops of Sec.~\ref{sec:cviolating} with non-identical inert scalars in the loops. For example, the magnitude of the CP-violating component of the $ZZZ$ vertex of diagram~{\footnotesize \circled{a}} in Fig.\ref{fig:ZZV} is proportional to the CP-violating weak-basis invariant~\cite{Lavoura:1994fv,Botella:1994cs}
\be
J_{CP} \propto \frac{(m_{S_i}^2-m_{S_j}^2)\:\!(m_{S_j}^2-m_{S_k}^2)\:\!(m_{S_k}^2-m_{S_i}^2)}{m_{S_i}^2m_{S_j}^2m_{S_k}^2} \,( g_{ZS_iS_j} )( g_{ZS_jS_k} )( g_{ZS_kS_i})\, ,
\label{eq:jcp}
\ee
which is non-zero only when the scalars $S_i,S_j,S_k$ are non-identical.
The C-odd P-even form factor $f_4$ for this process is given by a linear combination of the three-point tensor coefficient functions $C_{002}$ (in the notation of the \texttt{LoopTools} package~\cite{Hahn:2010zi}) and calculated to be~\cite{Cordero-Cid:2020yba}
\be
f_4^Z =
\frac{m_Z^2}{2\pi^2\, e \, (q_3^2 - m_Z^2)} \,
(g_{ZS_2S_3})( g_{ZS_1S_3} )( g_{ZS_1S_2})
\sum_{i,j,k=1}^3\epsilon_{ijk}C_{002}(m_Z^2,m_Z^2,q_3^2,m^2_i,m^2_j,m^2_k),
\label{eq:f4}
\ee
where $m_{i,j,k}$ stand for the mass of the $S_{i,j,k}$ scalars.

%%%%%%%%%%%%%%%%%%%%%%%%%%%%%%%%%%%%
%%%%%%%%%%%%%%%%%%%%%%%%%%%%%%%%%%%%
\section{Conclusions and outlook}
\label{sec:conclusions}
%%%%%%%%%%%%%%%%%%%%%%%%%%%%%%%%%%%%
%%%%%%%%%%%%%%%%%%%%%%%%%%%%%%%%%%%%

Electric dipole moments (EDMs) are among the most sensitive probes of new sources of CP violation, and their increasingly stringent experimental bounds place severe constraints on many baryogenesis-motivated extensions of the Standard Model (SM). In this work, we have identified a class of CP violation that is separated from these constraints by its discrete-symmetry properties, rather than only by small couplings or loop suppression. The key distinction lies in the discrete-symmetry structure: flavour-diagonal EDM operators are C-even and P-odd, whereas CP violation originating in a renormalisable scalar--gauge sector is C-odd and P-even. Indeed, the scalar and gauge interactions preserve parity, irrespective of the CP-violating scalar self-couplings, since they contain no parity-odd Lorentz structures. Consequently, CP violation confined to this sector cannot, by itself, generate the C-even, P-odd structure required for an EDM.

The I(2+1)HDM provides a concrete realisation of this mechanism. CP violation originates exclusively from the complex parameters of $V_{\Z_2}$ and remains confined to the inert scalar sector. The exact $\Z_2$ symmetry prevents mixing between the inert and active doublets and forbids Yukawa couplings of the inert scalars. As a result, the conventional one-loop and Barr--Zee contributions that dominate the electron EDM in CP-violating scalar extensions, such as the aligned 2HDM discussed in Appendix~B, are absent, independently of the inert scalar mass spectrum. The presence of two inert doublets is essential: a single inert doublet cannot accommodate a physical CP-violating phase, while removing the exact $\Z_2$ symmetry in a 2HDM generally allows CP violation to enter the fermionic sector. A dark sector containing one inert doublet and a complex singlet can reproduce part of the phenomenology, notably a reduced one-loop $WWZ$ contribution and the two-loop $WW\gamma$ contribution, whereas the $ZZZ$, $ZZ\gamma$ and $Z\gamma V$ vertices require two inert doublets. The I(2+1)HDM therefore represents the minimal realisation of the complete set of signatures investigated in this work.

Rather than manifesting itself through EDMs, this form of CP violation gives rise to a distinct set of observables. In triple gauge couplings, it contributes to the C-odd, P-even form factors $f_4^V$ and $h_{1,2}^V$, which vanish in the C-conserving limit. In the I(2+1)HDM, the $f_4^{WWZ}$ and $f_4^{ZZZ}$ form factors arise at one loop and are proportional to the CP-odd weak-basis invariants. By contrast, the $WW\gamma$, $ZZ\gamma$ and $Z\gamma V$ vertices first receive CP-violating contributions at two loops, as the diagonal nature of the photon couplings prevents the CP-violating phase from entering at one loop.
Complementary probes are provided by selection-rule tests involving decays of the same neutral scalar into final states of opposite C parity, such as $h\to ZV$ alongside $h\to ZZZ$ or $ZZ\gamma$, or $h\to\gamma\gamma$ alongside $h\to\gamma\gamma\gamma$, the latter arising only at three loops in the model. Since the SM already violates C, and generates the C-odd, P-odd structures $f_5^V$ and $h_{3,4}^V$, these tests point to the inert sector only once the P-even components are isolated and found to exceed the SM expectation. Static moments, on the other hand, do not provide equivalent tests: spin-zero inert scalars possess no static electromagnetic moments beyond their electric charge, while the C-transformation properties of fermionic moments do not, by themselves, reveal C violation in the underlying interactions.

An important qualification concerns the extent to which this symmetry-based separation protects the theory from EDM constraints. Within the scalar--gauge sector, parity conservation excludes the generation of an EDM operator, while the exact $\Z_2$ symmetry eliminates contributions requiring direct inert-scalar couplings to fermions. However, these properties do not imply that C and P are separately conserved in the full theory. The chiral gauge interactions of SM fermions violate both symmetries, and embedding a C-odd, P-even bosonic vertex into an amplitude involving non-conserved axial currents can, in principle, produce a C-even, P-odd contribution. For fermionic T-odd, P-even interactions such contributions are known to arise~\cite{Conti:1992xn,Ramsey-Musolf:1999cub,Kurylov:2000ub}. For the purely bosonic sources considered here, establishing whether they survive requires an analysis of the fully off-shell gauge-boson vertices and their embedding into complete gauge-invariant amplitudes; this is the content of our conjecture. In any case, since the P-odd bosonic operators of the SMEFT cannot be generated without a chiral fermion loop, any resulting EDM contribution arises only at higher loop order and carries additional electroweak suppression. A systematic investigation of these effects is underway.

Beyond their distinctive collider signatures, C-odd, P-even sources of CP violation are particularly interesting in the context of baryogenesis. The Sakharov conditions require both C and CP violation, making this class of interactions a natural candidate for generating the baryon asymmetry without relying on the conventional sources of CP violation constrained by EDM searches. Gauge interactions can provide a mechanism for communicating a dark-sector asymmetry to the visible sector, as demonstrated in a primordial setting in Ref.~\cite{Keus:2024khd}. Furthermore, the I(2+1)HDM can accommodate a strong first-order electroweak phase transition, providing another essential ingredient for electroweak baryogenesis~\cite{Keus:2025ova}. A dedicated investigation of the asymmetry-transfer mechanism and its cosmological implications will be presented separately.

As EDM experiments involving polar molecules, ultracold neutrons, diamagnetic atoms and storage rings continue to improve in sensitivity, the C-odd, P-even sources of CP violation identified here remain largely inaccessible to their conventional probes. Their phenomenology must instead be explored through complementary observables, particularly anomalous triple gauge couplings and C-sensitive scalar decays at the HL-LHC and future lepton colliders. The observation of such signatures, especially in the continued absence of an EDM signal, would provide compelling motivation for a qualitatively different origin of CP violation, one rooted in the symmetry structure of an extended scalar sector and potentially relevant to the generation of the baryon asymmetry of the Universe.

%%%%%%%%%%%%%%%%%%%%%%%%%%%%%%%%%%%%
%%%%%%%%%%%%%%%%%%%%%%%%%%%%%%%%%%%%
\subsection*{Acknowledgments}
The author is grateful to David~E.~Kaplan for numerous discussions and to Raman~Sundrum, Michael~Ramsey-Musolf and Anne-Katherine Burns for valuable comments. The author acknowledges financial support from Research Ireland grant 21/PATHS/9475 (MOREHIGGS) under the SFI-IRC Pathway Programme.

%%%%%%%%%%%%%%%%%%%%%%%%%%%%%%%%%%%%
%%%%%%%%%%%%%%%%%%%%%%%%%%%%%%%%%%%%
%%%%%%%%%%%%%%%%%%%%%%%%%%%%%%%%%%%%

%%%%%%%%%%%%%%%%%%%%%%%%%%%%%%%%%%%%
%%%%%%%%%%%%%%%%%%%%%%%%%%%%%%%%%%%%
\begin{appendices}
%%%%%%%%%%%%%%%%%%%%%%%%%%%%%%%%%%%%
%%%%%%%%%%%%%%%%%%%%%%%%%%%%%%%%%%%%

%%%%%%%%%%%%%%%%%%%%%%%%%%%%%%%%%%%%
%%%%%%%%%%%%%%%%%%%%%%%%%%%%%%%%%%%%
\section{Calculation of electron EDM contributions}
\label{app:EDMs}
%%%%%%%%%%%%%%%%%%%%%%%%%%%%%%%%%%%%
%%%%%%%%%%%%%%%%%%%%%%%%%%%%%%%%%%%%

The electron
electric dipole moment (eEDM), $d_e$, is defined through the effective Lagrangian
\begin{align}
\mathcal{L}_{d_e} &= -\dfrac{i}{2}\:\!d_e\:\!(\bar{e}\sigma^{\mu\nu}\gamma_5 e)\:\!F_{\mu\nu}\:\!,
\label{eq:de}
\end{align}
where $F_{\mu\nu}$ is the electromagnetic field strength tensor and
$\sigma^{\mu\nu} = \nicefrac{i}{2}[\gamma^\mu, \gamma^\nu]$. The relevant part of the Lagrangian
mediating these interactions can be written as
\be
\mathcal{L} \supset \frac{e\:\! m_e}{8\pi^2}
\left[c_L\:\!(\bar{e}\sigma^{\mu\nu}P_L e)\:\!F_{\mu\nu}
+ c_R\:\!(\bar{e}\sigma^{\mu\nu}P_R e)\:\!F_{\mu\nu} + \text{h.c.}\right],
\label{eq:Leff}
\ee
where $P_{L,R} = \nicefrac{1}{2}(1 \mp \gamma_5)$ are the left-
and right-handed projection operators. Expanding Eq.(\ref{eq:Leff}) explicitly gives
\be
\mathcal{L} \supset \frac{e\:\! m_e}{8\pi^2}\:\!\text{Re}(c_L + c^*_R)\:\!
(\bar{e}\sigma^{\mu\nu}e)\:\!F_{\mu\nu}
- \frac{i\:\!e\:\! m_e}{8\pi^2}\:\!\text{Im}(c_L + c^*_R)\:\!
(\bar{e}\sigma^{\mu\nu}\gamma_5 e)\:\!F_{\mu\nu}\:\!,
\label{eq:Lexpanded}
\ee
from which one can identify
\be
d_e = \frac{e\:\! m_e}{4\pi^2}\:\!\text{Im}(c_L + c^*_R)\,.
\label{eq:deWilson}
\ee
The Wilson coefficients $c_L$ and $c_R$ are calculated separately for each loop diagram
contributing to these observables.

The ordinary CP-violating 2HDM with a softly broken $\mathbb{Z}_2$ symmetry~\cite{Keus:2015hva,Keus:2017ioh} (or similar scalar extensions) possesses scalar-fermion couplings already at tree level, which consequently contribute to EDMs through the usual one- and two-loop Barr-Zee diagrams as shown below.

\subsection{One-loop contributions of scalars}
\label{sec:oneloop}

\begin{figure}[h!]
\begin{minipage}{\linewidth}
\begin{center}
\begin{tikzpicture}[thick,scale=1.0]
\draw (0,0) -- node[black,above,yshift=-0.8cm,xshift=0.0cm] {$ $} (4,0);
\draw[particle] (0,0) -- node[black,above,sloped,yshift=-0.3cm,xshift=-0.9cm] {$e$} (1,0);
\draw[particle] (1,0) -- node[black,above,sloped,yshift=-0.0cm,xshift=0.0cm] {$e$} (2,0);
\draw[particle] (3,0) -- node[black,above,sloped,yshift=-0.3cm,xshift=0.9cm] {$e$} (4,0);
\draw[photon] (2,0) -- node[black,above,yshift=-0.4cm,xshift=0.5cm] {$\gamma$} (3,-1);
\draw[dashed]  (1,0) node[black,above,sloped,yshift=0.95cm,xshift=1.05cm] {$h_i$}  arc (180:0:1cm) ;
\draw[particle] (2,0) -- node[black,above,sloped,yshift=-0.0cm,xshift=0.0cm] {$ e$} (3,0);
\end{tikzpicture}
\end{center}
\vspace{-5mm}
\caption{\small The 1-loop diagram mediated by neutral scalars $h_i$ contributing to $e$EDM.}
\label{fig:oneloop}
\end{minipage}
\end{figure}
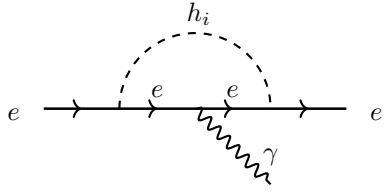

The one-loop diagram contributing to $d_e$ is shown in
Fig.\ref{fig:oneloop}, where $h_i$ are the neutral scalars of the model with couplings
to electrons denoted by $Y^{h_i}_{ee}$.
The charged scalar mediated version of this diagram is sub-dominant and is therefore
neglected~\cite{Chun:2015xfx}.
The one-loop contributions to $d_e$ are
\begin{align}
d^{1\text{-loop}}_e &= \frac{e\:\! m_e}{16\pi^2}
\sum^n_{i=1}\:\!\text{Im}\!\left[(Y^{h_i}_{ee})^2\right]
\int^1_0 dx \int^x_0 dy\:\!
\frac{(y-1)}
{m^2_e\left[y(y-x) + (1-y)\right] + m^2_{h_i}\:\!y}\:\!,
\label{eq:de1loop}
\end{align}
where $n$ is the number of neutral scalars mediating the loop in Fig.\ref{fig:oneloop}.
\vspace{0.2cm}

\subsection{Two-loop contributions of scalars}
\label{sec:twoloop}

\begin{figure}[h]
\begin{minipage}{\linewidth}
\centering
\begin{minipage}{\linewidth}
\begin{tikzpicture}[thick,scale=1.0]
\draw[particle] (-0.5,0) -- node[black,above,sloped,yshift=-0.0cm,xshift=-0.0cm] {$e$} (0.5,0);
\draw[particle] (0.5,0) -- node[black,above,sloped,yshift=-0.0cm,xshift=0.0cm] {$e$} (3.5,0);
\draw[particle] (3.5,0) -- node[black,above,sloped,yshift=-0.0cm,xshift=0.0cm] {$e$} (4.5,0);
\draw[dashed] (0.5,0) -- node[black,above,yshift=-0.5cm,xshift=-0.3cm] {$h_i$} (1.293,-1.293);
\draw[photon] (3.5,0) -- node[black,above,yshift=-0.4cm,xshift=0.5cm] {$\gamma,Z$} (2.707,-1.293);
\draw[photon] (2,-3) -- node[black,above,yshift=-0.6cm,xshift=0.3cm] {$\gamma$} (2,-4);
\draw[particle]  (2,-1) node[black,above,sloped,yshift=-1.2cm,xshift=-1.2cm] {$f$}  arc (90:270:1cm) ;
\draw[particle]  (2,-3) node[black,above,sloped,yshift=0.8cm,xshift=1.2cm] {$f$}  arc (-90:90:1cm) ;
\end{tikzpicture}
%%%%%%%%%%%%%%%%%%%%%%%%%%%%%
\hspace{2mm}
%%%%%%%%%%%%%%%%%%%%%
\begin{tikzpicture}[thick,scale=1.0]
\draw[particle] (-0.5,0) -- node[black,above,sloped,yshift=-0.0cm,xshift=-0.0cm] {$e$} (0.5,0);
\draw[particle] (0.5,0) -- node[black,above,sloped,yshift=-0.0cm,xshift=0.0cm] {$e$} (3.5,0);
\draw[particle] (3.5,0) -- node[black,above,sloped,yshift=-0.0cm,xshift=0.0cm] {$e$} (4.5,0);
\draw[dashed] (0.5,0) -- node[black,above,yshift=-0.5cm,xshift=-0.3cm] {$h_i$} (1.293,-1.293);
\draw[photon] (3.5,0) -- node[black,above,yshift=-0.4cm,xshift=0.5cm] {$\gamma,Z$} (2.707,-1.293);
\draw[photon] (2,-3) -- node[black,above,yshift=-0.6cm,xshift=0.3cm] {$\gamma$} (2,-4);
\draw[photon]  (2.707,-1.293) node[black,above,sloped,yshift=-1.2cm,xshift=-2.1cm] {$W$}  arc (45:405:1cm) ;
%\end{tikzpicture}
%%%%%%%%%%%%%%%%%%%%%%%%%%%%%
\hspace{55mm}
%%%%%%%%%%%%%%%%%%%%%
%\begin{tikzpicture}[thick,scale=1.0]
\draw[particle] (-0.5,0) -- node[black,above,sloped,yshift=-0.0cm,xshift=-0.0cm] {$e$} (0.5,0);
\draw[particle] (0.5,0) -- node[black,above,sloped,yshift=-0.0cm,xshift=0.0cm] {$e$} (3.5,0);
\draw[particle] (3.5,0) -- node[black,above,sloped,yshift=-0.0cm,xshift=0.0cm] {$e$} (4.5,0);
\draw[dashed] (0.5,0) -- node[black,above,yshift=-0.5cm,xshift=-0.3cm] {$h_i$} (1.293,-1.293);
\draw[photon] (3.5,0) -- node[black,above,yshift=-0.4cm,xshift=0.5cm] {$\gamma,Z$} (3,-2);
\draw[photon] (3,-2) -- node[black,above,yshift=-0.6cm,xshift=0.7cm] {$\gamma$} (4,-3);
\draw[photon]  (3,-2) node[black,above,sloped,yshift=-1.2cm,xshift=-2.1cm] {$W$}  arc (0:360:1cm) ;
\end{tikzpicture}
\vspace{-1mm}
\caption{\small The Barr--Zee diagrams with largest contributions to eEDM.}
\label{fig:BarrZee}
\end{minipage}
\end{minipage}
\end{figure}
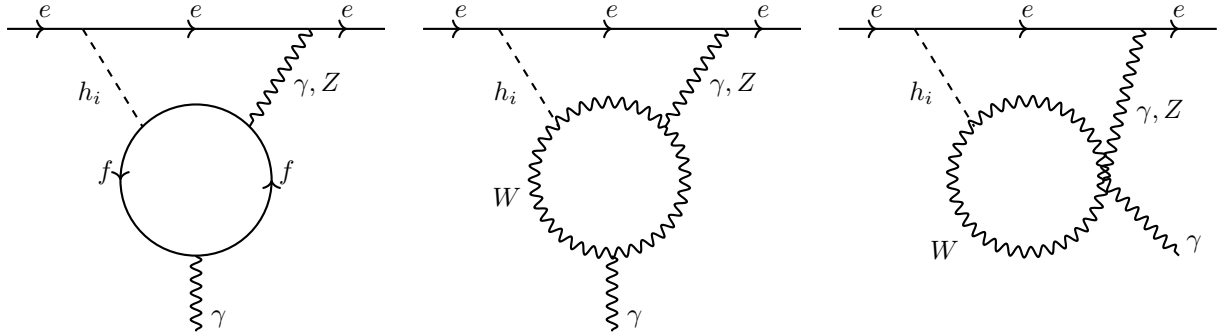

The dominant two-loop contributions to $d_e$ arise from the Barr--Zee
diagrams mediated by the scalar states, shown in Fig.\ref{fig:BarrZee}. Diagrams with
a $Z$ boson in the loop (instead of a photon) are suppressed by a factor of
$(\frac{1}{4} - \sin^2\theta_W)$, making their contribution almost two orders of
magnitude smaller than the photon-mediated diagrams; these are therefore neglected.
Similarly, contributions from charged scalars are sub-dominant and are also
neglected~\cite{Cheung:2001hz}. For the two-loop calculations, we use the results of
Ref.~\cite{Harnik:2012pb}.

The two-loop contributions to $d_e$ from heavy fermion loops ($f = t, b, c, \tau$)
and $W$ boson loops are\footnote{The subscript $f$ here denotes a fermion and is not to
be confused with the loop function $f(z)$.}
\begin{align}
d^{2\text{-loop}}_{e,\:\!f} &= \frac{e}{3\pi^2}
\left(\frac{\alpha\:\! G_F\:\! v^2}{\sqrt{2}\:\!\pi\:\! m_f}\right)
\sum^n_{i=1}\left[\text{Im}(Y^{h_i}_{ee})\:\!\text{Re}(Y^{h_i}_{ff})\:\!f(z_{fh_i})
+ \text{Re}(Y^{h_i}_{ee})\:\!\text{Im}(Y^{h_i}_{ff})\:\!g(z_{fh_i})\right],
\label{eq:de2loopf} \\[6pt]
d^{2\text{-loop}}_{e,\:\!W} &= -\frac{e}{8\pi^2}
\left(\frac{\alpha\:\! G_F\:\! v}{\sqrt{2}\:\!\pi}\right)
\sum^n_{i=1} \frac{Y^{h_i}_{WW}}{2m^2_W/v}\:\!\text{Im}(Y^{h_i}_{ee})
\nonumber \\
&\quad \times \left[3f(z_{Wh_i}) + \frac{23}{4}g(z_{Wh_i}) + \frac{3}{4}h(z_{Wh_i})
+ \frac{f(z_{Wh_i}) - g(z_{Wh_i})}{2z_{Wh_i}}\right].
\label{eq:de2loopW}
\end{align}
where $z_{AB} = m^2_A/m^2_B$ and $Y^{h_i}_{WW}$ is the coupling of $h_i$ to $WW$. For
the SM-Higgs coupling to $WW$, we use the notation $Y^{h_\text{SM}}_{WW} = 2m^2_W/v$.
The loop functions $f(z)$, $g(z)$ and $h(z)$ are:
\bea
f(z)&=&\frac{1}{2}z\int_{0}^{1}dx\frac{1-2x(1-x)}{x(1-x)-z}\log\left(\frac{x(1-x)}{z}\right),\\
g(z)&=&\frac{1}{2}z\int_{0}^{1}dx\frac{1}{x(1-x)-z}\log\left(\frac{x(1-x)}{z}\right),\\
h(z)&=&z^{2}\frac{\partial}{\partial z}\left(\frac{g(z)}{z}\right)=\frac{z}{2}\int_{0}^{1}\frac{dx}{z-x(1-x)}\left[1+\frac{z}{z-x(1-x)}\log\left(\frac{x(1-x)}{z}\right)\right].
\eea

Therefore, the couplings that are important in EDM contributions for scalar extensions with CP violation are $Y^{h_i}_{ee}$, $Y^{h_i}_{ff}$ and $Y^{h_i}_{WW}$.

%%%%%%%%%%%%%%%%%%%%%%
\subsection{Zero EDM contributions from dark CP violation}
\label{sec:zeroEDM}
%%%%%%%%%%%%%%%%%%%%%%

The $e$EDM receives contributions only through the Yukawa couplings $Y^{h_i}_{ee}$ of the
neutral scalars to the electron, which in turn depend on the mixing between the scalar
mass eigenstates and the active doublet $\phi_3$. In the I(2+1)HDM, the exact $\Z_2$
symmetry forbids any mixing between the inert scalars $S_{1,2,3,4}$ and the active
sector. Consequently, the inert scalars have no Yukawa couplings to SM fermions,
\be
Y^{S_i}_{ee} =Y^{S_i}_{ff} = 0 \qquad
\text{for all inert scalars } S_i\:\!,
\label{eq:zeroYukawa}
\ee
and the sums in Eqs.(\ref{eq:de1loop})--(\ref{eq:de2loopW}) receive contributions only
from the SM-like Higgs boson $h$, whose couplings are CP-conserving. It therefore follows directly that $d_e = 0$ in all these diagrams. Contributions in which the inert-sector CP violation reaches the electron only through gauge-boson vertices are not of this type, and are the subject of the conjecture in Sec.~\ref{sec:2V*-cancellation}.

%%%%%%%%%%%%%%%%%%%%%%%%%%%%%%%%%%%%
\section{C-violating CP violation leakage in 2HDM}
\label{app:2hdm}
%%%%%%%%%%%%%%%%%%%%%%%%%%%%%%%%%%%%

The EDM contributions of dark CP violation through scalar--fermion couplings vanish exactly, as a direct consequence of the $\mathbb{Z}_2$ symmetry that forbids Yukawa couplings of the inert scalars. This is in sharp contrast to the
CP-violating 2HDM where the absence of such protecting symmetry allows CP violation to seep into the fermionic sector, as we show next.

In the case of the \textit{flavour-aligned} 2HDM, it is possible to have such CP-violating sources due to the absence of the $\Z_2$ symmetry.
In terms of the parameters of the potential in the so-called \textit{Higgs basis}~\cite{Georgi:1978ri,Lavoura:1994yu,Lavoura:1994fv,Botella:1994cs,Branco:1999fs,Davidson:2005cw}, this leads to $Z_7$ being the only complex parameter in the potential,  in the exact \textit{Higgs-alignment limit}~\cite{Ginzburg:2001wj,Gunion:2002zf,Craig:2013hca,Asner:2013psa,Carena:2013ooa,Haber:2013mia}.
As a result, the interactions of $h_1$ remain CP-conserving in the exact Higgs alignment limit, while cubic and quartic
scalar interactions involving $h_2$ and $h_3$ violate CP, such as in the $h_1 h_2 h_3$ vertex~\cite{Haber:2022gsn}.

%The Higgs-basis scalar potential is given by,
%\bea
%\mathcal{V}&=& Y_1 \widehat{\Phi}_1^\dagger \widehat{\Phi}_1+ Y_2 \widehat{\Phi}_2^\dagger \widehat{\Phi}_2 +[Y_3 \widehat{\Phi}_1^\dagger \widehat{\Phi}_2+{\rm h.c.}]
%\nn\\
%&&\quad
%+\half Z_1(\widehat{\Phi}_1^\dagger \widehat{\Phi}_1)^2+\half Z_2(\widehat{\Phi}_2^\dagger \widehat{\Phi}_2)^2
%+Z_3(\widehat{\Phi}_1^\dagger \widehat{\Phi}_1)(\widehat{\Phi}_2^\dagger \widehat{\Phi}_2)
%+Z_4(\widehat{\Phi}_1^\dagger \widehat{\Phi}_2)(\widehat{\Phi}_2^\dagger \widehat{\Phi}_1) \nn \\
%&&\quad
%+\left\{\half Z_5 (\widehat{\Phi}_1^\dagger \widehat{\Phi}_2)^2 +\big[Z_6  (\widehat{\Phi}_1^\dagger
%\widehat{\Phi}_1) +Z_7  (\widehat{\Phi}_2^\dagger \widehat{\Phi}_2)\big] \widehat{\Phi}_1^\dagger \widehat{\Phi}_2+{\rm
%h.c.}\right\}.
%\label{higgspot}
%\eea

\begin{figure}[h!]
\begin{minipage}{\linewidth}
\begin{center}
\begin{tikzpicture}[thick,scale=1.0]
\draw[particle] (0,0) -- node[black,above,sloped,yshift=-0.3cm,xshift=-0.9cm] {$e$} (1,0);
\draw[particle] (1,0) -- node[black,above,sloped,yshift=-0.5cm,xshift=0.0cm] {$e$} (2,0);
\draw[particle] (2,0) -- node[black,above,sloped,yshift=-0.5cm,xshift=0.1cm] {$e$} (3,0);
\draw[particle] (3,0) -- node[black,above,sloped,yshift=-0.3cm,xshift=0.9cm] {$e$} (4,0);
% Photon
\draw[photon]
(2,0) -- node[black,above,yshift=-0.7cm,xshift=-0.1cm] {$\gamma$} (3,-1.2);
% Left dashed Higgs propagator: from (1,0) up to (1.4,1.2)
\draw[dashed] (1,0) -- node[black,above,yshift=-0.2cm,xshift=-0.3cm] {$h_1$} (1.4,1.2);
% Right dashed Higgs propagator: from (2.6,1.2) down to (3,0)
\draw[dashed] (2.6,1.1) -- node[black,above,yshift=-0.3cm,xshift=0.3cm] {$h_1$} (3,-0.1);
% Scalar loop bubble centered at (2, 1.2) with radius 0.6cm
\draw[dashed] (1.4,1.2) arc (180:540:0.6cm);
%\draw[dashed] (1.4,1.2) arc (180:360:0.6cm);
% Scalar label inside the loop
\node at (2,2.1) {$h_{2}$};
\node at (2,1.0) {$h_{3}$};
\end{tikzpicture}
\vspace{-1mm}
\caption{\small The 2-loop extension of the lowest order diagrams contributing to the electron EDM in the CP-violating aligned 2HDM.}
\label{fig:BarrZee-new1}
\end{center}
\end{minipage}
\end{figure}
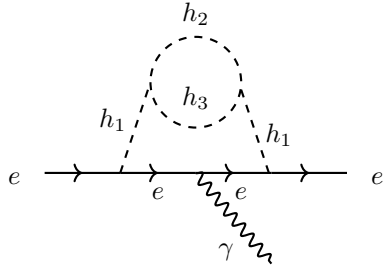

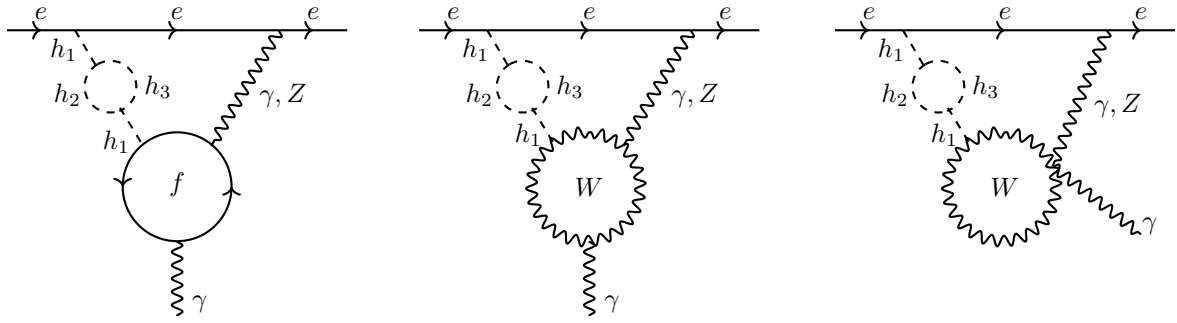
\begin{figure}[h!]
\begin{center}
\begin{minipage}{\linewidth}
\begin{minipage}{0.3\linewidth}
\begin{tikzpicture}[thick,scale=0.9]
% Electron line
\draw[particle] (-0.5,0) -- node[black,above,sloped,yshift=-0.0cm,xshift=-0.0cm] {$e$} (0.5,0);
\draw[particle] (0.5,0) -- node[black,above,sloped,yshift=-0.0cm,xshift=0.0cm] {$e$} (3.5,0);
\draw[particle] (3.5,0) -- node[black,above,sloped,yshift=-0.0cm,xshift=0.0cm] {$e$} (4.5,0);
\draw[dashed] (0.5,0) -- node[black,above,yshift=-0.3cm,xshift=-0.3cm] {$h_1$} (0.83,-0.56);% (0.7,-0.5);
\draw[dashed] (1.16,-1.14) -- node[black,above,yshift=-0.5cm,xshift=-0.2cm] {$h_1$} (1.5,-1.7);% (0.7,-0.5);
%\draw[dashed] (1.166,-0.916) -- (1.293,-1.7);
\draw[dashed] (0.83,-0.5) node[black,above,yshift=-0.7cm,xshift=-0.4cm] {$h_2$} arc (120:-70:0.38cm);
\draw[dashed] (0.83,-0.5) node[black,above,yshift=-0.6cm,xshift=0.8cm] {$h_3$} arc (120:290:0.38cm);
\draw[photon](3.5,0) -- node[black,above,yshift=-0.4cm,xshift=0.5cm] {$\gamma,Z$} (2.5,-1.7);
\draw[photon](2,-3.1) -- node[black,above,yshift=-0.6cm,xshift=0.3cm] {$\gamma$} (2,-4.2);
\draw[particle] (2,-1.5) node[black,above,sloped,yshift=-1.0cm,xshift=-0.0cm] {$ f$} arc (90:270:0.8cm);
\draw[antiparticle] (2,-1.5) node[black,above,sloped,yshift=0.8cm,xshift=1.2cm]  {$ $} arc (90:-90:0.8cm);
\end{tikzpicture}
\end{minipage}
%%%%%%%%%%%%%%%%%%%%%%%%%%%%%
\hspace{2mm}
%%%%%%%%%%%%%%%%%%%%%
\begin{minipage}{0.3\linewidth}
\begin{tikzpicture}[thick,scale=0.9]
\draw[particle] (-0.5,0) -- node[black,above,sloped,yshift=-0.0cm,xshift=-0.0cm] {$e$} (0.5,0);
\draw[particle] (0.5,0) -- node[black,above,sloped,yshift=-0.0cm,xshift=0.0cm] {$e$} (3.5,0);
\draw[particle] (3.5,0) -- node[black,above,sloped,yshift=-0.0cm,xshift=0.0cm] {$e$} (4.5,0);
\draw[dashed] (0.5,0) -- node[black,above,yshift=-0.3cm,xshift=-0.3cm] {$h_1$} (0.83,-0.56);% (0.7,-0.5);
\draw[dashed] (1.16,-1.14) -- node[black,above,yshift=-0.4cm,xshift=-0.2cm] {$h_1$} (1.5,-1.7);% (0.7,-0.5);
%\draw[dashed] (1.166,-0.916) -- (1.293,-1.7);
\draw[dashed] (0.83,-0.5) node[black,above,yshift=-0.7cm,xshift=-0.4cm] {$h_2$} arc (120:-70:0.38cm);
\draw[dashed] (0.83,-0.5) node[black,above,yshift=-0.6cm,xshift=0.8cm] {$h_3$} arc (120:290:0.38cm);
\draw[photon](3.5,0) -- node[black,above,yshift=-0.4cm,xshift=0.5cm] {$\gamma,Z$} (2.5,-1.7);
\draw[photon](2,-3.1) -- node[black,above,yshift=-0.6cm,xshift=0.3cm] {$\gamma$} (2,-4.2);
\draw[photon] (2,-1.5) node[black,above,sloped,yshift=-1.0cm,xshift=-0.0cm] {$ W $} arc (90:450:0.8cm);
%\draw[decorate,decoration={snake,amplitude=2pt,segment length=5pt},black] (2,-1.5) node[black,above,sloped,yshift=0.8cm,xshift=1.2cm]  {$ $} arc (90:-90:0.8cm);
%\end{tikzpicture}
%\end{minipage}
%%%%%%%%%%%%%%%%%%%%%%%%%%%%%
\hspace{55mm}
%%%%%%%%%%%%%%%%%%%%%
%\begin{minipage}{0.3\linewidth}
%\begin{tikzpicture}[thick,scale=0.9]
% Electron line
\draw[particle] (-0.5,0) -- node[black,above,sloped,yshift=-0.0cm,xshift=-0.0cm] {$e$} (0.5,0);
\draw[particle] (0.5,0) -- node[black,above,sloped,yshift=-0.0cm,xshift=0.0cm] {$e$} (3.5,0);
\draw[particle] (3.5,0) -- node[black,above,sloped,yshift=-0.0cm,xshift=0.0cm] {$e$} (4.5,0);
\draw[dashed] (0.5,0) -- node[black,above,yshift=-0.3cm,xshift=-0.3cm] {$h_1$} (0.83,-0.56);% (0.7,-0.5);
\draw[dashed] (1.16,-1.14) -- node[black,above,yshift=-0.4cm,xshift=-0.2cm] {$h_1$} (1.5,-1.7);% (0.7,-0.5);
%\draw[dashed] (1.166,-0.916) -- (1.293,-1.7);
\draw[dashed] (0.83,-0.5) node[black,above,yshift=-0.7cm,xshift=-0.4cm] {$h_2$} arc (120:-70:0.38cm);
\draw[dashed] (0.83,-0.5) node[black,above,yshift=-0.6cm,xshift=0.8cm] {$h_3$} arc (120:290:0.38cm);
\draw[photon] (3.5,0) -- node[black,above,yshift=-0.4cm,xshift=0.5cm] {$\gamma,Z$} (2.7,-2);
\draw[photon] (2.7,-2) -- node[black,above,yshift=-0.6cm,xshift=0.7cm] {$\gamma$} (4,-3);
\draw[photon] (2,-1.5) node[black,above,sloped,yshift=-1.0cm,xshift=-0.0cm] {$ W $} arc (90:450:0.8cm);
\end{tikzpicture}
\end{minipage}
\vspace{-1mm}
\caption{\small The 3-loop extension of the Barr--Zee-type diagrams contributing to the electron EDM in the CP-violating aligned 2HDM.}
\label{fig:BarrZee-new2}
\end{minipage}
\end{center}
\end{figure}

Such a source of CP violation does contribute to the electron EDM, at three loops through a scalar loop inserted into the Barr--Zee diagrams, and at two loops through a scalar loop inserted into the one-loop scalar-exchange diagram, whose contribution is strongly suppressed by the squared electron Yukawa coupling $Y_e^2$, as shown in Figs.\ref{fig:BarrZee-new1}--\ref{fig:BarrZee-new2}.

%%%%%%%%%%%%%%%%%%%%%%%%%%%%%%%%%%%%
%%%%%%%%%%%%%%%%%%%%%%%%%%%%%%%%%%%%
\section{The effective operators for $V_1^\ast V_2^\ast V_3^\ast $}
\label{app:fully-off-shell}
%%%%%%%%%%%%%%%%%%%%%%%%%%%%%%%%%%%%
%%%%%%%%%%%%%%%%%%%%%%%%%%%%%%%%%%%%

The effective operators for the $V_1^\ast V_2^\ast V_3^\ast $ process with all three vector bosons off-shell are given in Ref.~\cite{Gounaris:2000dn,Degrande:2013kka,Gavela:2014vra,Subba:2023rpm,Ellis:2023ucy,Ellis:2024omd}. For the reader's convenience, we show below the CP-violating operators for the $Z^*Z^*Z^*$, $Z^*Z^*\gamma^*$ , $\gamma^*\gamma^*Z^*$ and $W^*W^*V^*$ processes in Tabs.~\ref{tab:ZZZ-CPv-op}, \ref{tab:ZZgamma-CPv-op}, \ref{tab:gammagammaZ-CPv-op} and \ref{tab:WWV-CPv-op}, respectively.

Note that in order to generate all independent Lorentz structures of the fully off-shell vertices, we have added the minimal set of higher-dimensional operators.

\begin{footnotesize}
\begin{table}[htbp]
\centering
\renewcommand{\arraystretch}{1.3}
\begin{tabular}{|c| p{0.40\textwidth}| c|}
\hline
\textbf{Operator} & \textbf{Expression} & \textbf{Dimension} \\
\hline
$\widetilde{\O}^{Z^*Z^*Z^*}_{1}$ &
$-Z_{\sigma}(\partial^{\sigma}Z_{\nu})(\partial_{\mu}Z^{\mu\nu})$ & dim=6
\\
\hline
$\widetilde{\O}^{Z^*Z^*Z^*}_{2}$ &
$(\square Z_{\alpha})(\partial^{\alpha}Z_{\mu})(\square Z^{\mu})$ & dim=8
\\
\hline
$\widetilde{\O}^{Z^*Z^*Z^*}_{3}$ &
$Z_{\alpha}(\partial^{\alpha}Z_{\mu})(\square^2 Z^{\mu})$ & dim=8
\\
\hline
$\widetilde{\O}^{Z^*Z^*Z^*}_{4}$ &
$(\square^2\partial^{\alpha}Z_{\beta})
(\partial^{\mu}\square Z_{\alpha})
(\partial^{\beta}Z_{\mu})$ & dim=12
\\
\hline
$\widetilde{\O}^{Z^*Z^*Z^*}_{5}$ &
$Z^{\mu}Z_{\mu}(\partial^{\sigma}Z_{\sigma})$ & dim=4
\\
\hline
$\widetilde{\O}^{Z^*Z^*Z^*}_{6}$ &
$(\square Z^{\mu})Z_{\mu}(\partial^{\sigma}Z_{\sigma})$ & dim=6
\\
\hline
$\widetilde{\O}^{Z^*Z^*Z^*}_{7}$ &
$Z^{\mu}Z_{\mu}\square(\partial^{\sigma}Z_{\sigma})$ & dim=6
\\
\hline
$\widetilde{\O}^{Z^*Z^*Z^*}_{8}$ &
$(\partial^{\sigma}Z_{\sigma})(\partial^{\nu}Z_{\mu})
(\partial^{\mu}Z_{\nu})$ & dim=6
\\
\hline
$\widetilde{\O}^{Z^*Z^*Z^*}_{9}$ &
$(\partial^{\sigma}Z_{\sigma})(\square\partial^{\alpha}Z_{\beta})
(\partial^{\beta}Z_{\alpha})$ & dim=8
\\
\hline
$\widetilde{\O}^{Z^*Z^*Z^*}_{10}$ &
$(\square\partial^{\sigma}Z_{\sigma})
(\partial^{\alpha}Z_{\beta})(\partial^{\beta}Z_{\alpha})$ & dim=8
\\
\hline
$\widetilde{\O}^{Z^*Z^*Z^*}_{11}$ &
$\square\partial^{\alpha}(\partial^{\sigma}Z_{\sigma})
(\partial^{\beta}Z_{\beta})Z_{\alpha}$ & dim=8
\\
\hline
$\widetilde{\O}^{Z^*Z^*Z^*}_{12}$ &
$\square\partial^{\alpha}(\partial^{\sigma}Z_{\sigma})
(\partial^{\beta}Z_{\beta})(\square Z_{\alpha})$ & dim=10
\\
\hline
$\widetilde{\O}^{Z^*Z^*Z^*}_{13}$ &
$\square^2\partial^{\alpha}(\partial^{\sigma}Z_{\sigma})
(\partial^{\beta}Z_{\beta})Z_{\alpha}$ & dim=10
\\
\hline
$\widetilde{\O}^{Z^*Z^*Z^*}_{14}$ &
$(\partial^{\sigma}Z_{\sigma})(\partial^{\mu}Z_{\mu})
(\partial^{\nu}Z_{\nu})$ & dim=6
\\
\hline
\end{tabular}
\caption{The CP-violating operators for the $Z^*Z^*Z^*$ vertex}
\label{tab:ZZZ-CPv-op}
\end{table}
\end{footnotesize}

\begin{footnotesize}
\begin{table}[htbp]
\centering
\renewcommand{\arraystretch}{1.3}
\begin{tabular}{|c| p{0.40\textwidth}| c|}
\hline
\textbf{Operator} & \textbf{Expression} & \textbf{Dimension} \\
\hline
$\widetilde{\O}^{Z^*Z^*\gamma^*}_{1}$ &
$-F^{\mu\beta} Z_{\beta} (\partial^{\sigma}Z_{\sigma\mu})$ & dim=6
\\
\hline
$\widetilde{\O}^{Z^*Z^*\gamma^*}_{2}$ &
$-(\partial_{\alpha}\partial_{\beta}\square Z_{\mu})
Z^{\alpha}F^{\mu\beta}$ & dim=8
\\
\hline
$\widetilde{\O}^{Z^*Z^*\gamma^*}_{3}$ &
$-(\partial_{\mu}F^{\mu\beta})Z_{\alpha}(\partial^{\alpha}Z_{\beta})$ & dim=6
\\
\hline
$\widetilde{\O}^{Z^*Z^*\gamma^*}_{4}$ &
$\partial^{\mu}F_{\mu\nu}(\square\partial^{\nu}Z_{\alpha})Z^{\alpha}$ & dim=8
\\
\hline
$\widetilde{\O}^{Z^*Z^*\gamma^*}_{5}$ &
$(\partial^{\sigma}Z_{\sigma})F_{\mu\nu}(\partial^{\mu}Z^{\nu})$ & dim=6
\\
\hline
$\widetilde{\O}^{Z^*Z^*\gamma^*}_{6}$ &
$(\partial^{\sigma}Z_{\sigma})(\partial^{\mu}F_{\mu\nu})Z^{\nu}$ & dim=6
\\
\hline
$\widetilde{\O}^{Z^*Z^*\gamma^*}_{7}$ &
$\square(\partial^{\sigma}Z_{\sigma})F_{\mu\nu}(\partial^{\mu}Z^{\nu})$ & dim=8
\\
\hline
$\widetilde{\O}^{Z^*Z^*\gamma^*}_{8}$ &
$(\partial^{\sigma}Z_{\sigma})F_{\mu\nu}(\square\partial^{\mu}Z^{\nu})$ & dim=8
\\
\hline
$\widetilde{\O}^{Z^*Z^*\gamma^*}_{9}$ &
$\square\partial^{\nu}(\partial^{\sigma}Z_{\sigma})(\partial^{\beta}
Z_{\beta})\partial^{\mu}F_{\mu\nu}$ & dim=10
\\
\hline
\end{tabular}
\caption{The CP-violating operators for the $Z^*Z^*\gamma^*$ vertex}
\label{tab:ZZgamma-CPv-op}
\end{table}
\end{footnotesize}

\begin{footnotesize}
\begin{table}[htbp]
\centering
\renewcommand{\arraystretch}{1.3}
\begin{tabular}{|c| p{0.40\textwidth}| c|}
\hline
\textbf{Operator} & \textbf{Expression} & \textbf{Dimension} \\
\hline
$\widetilde{\O}^{\gamma^*\gamma^*Z^*}_{1}$ &
$-(\partial^{\sigma}F_{\sigma\mu})Z_{\beta}F^{\mu\beta}$ & dim=6
\\
\hline
$\widetilde{\O}^{\gamma^*\gamma^*Z^*}_{2}$ &
$(\square F^{\mu\nu})F_{\nu\alpha}(\partial_{\mu}Z^{\alpha})$ & dim=8
\\
\hline
$\widetilde{\O}^{\gamma^*\gamma^*Z^*}_{3}$ &
$-(\partial_{\alpha}\partial_{\beta}\partial^{\rho}F_{\rho\mu})Z^{\alpha}
F^{\mu\beta}$ & dim=8
\\
\hline
$\widetilde{\O}^{\gamma^*\gamma^*Z^*}_{4}$ &
$(\square \partial_{\mu}F^{\mu\nu})
(\partial^{\sigma}F_{\sigma\alpha})(\partial_{\nu}Z^{\alpha})$ & dim=10
\\
\hline
$\widetilde{\O}^{\gamma^*\gamma^*Z^*}_{5}$ &
$(\partial_{\sigma}Z^{\sigma})F^{\mu\nu}F_{\mu\nu}$ & dim=6
\\
\hline
$\widetilde{\O}^{\gamma^*\gamma^*Z^*}_{6}$ &
$\partial_{\mu}(\partial_{\sigma}Z^{\sigma})F^{\mu\nu}
(\partial^{\beta}F_{\beta\nu})$ & dim=8
\\
\hline
\end{tabular}
\caption{The CP-violating operators for the $\gamma^*\gamma^*Z^*$ vertex}
\label{tab:gammagammaZ-CPv-op}
\end{table}
\end{footnotesize}

\begin{footnotesize}
\begin{table}[htbp]
\centering
\renewcommand{\arraystretch}{1.3}
\begin{tabular}{|c| p{0.4\textwidth}| c|}
\hline
\textbf{Operator} & \textbf{Expression} & \textbf{Dimension} \\
\hline
$\widetilde{\O}^{W^{+*}W^{-*}V^*}_{1}$ &
$W^-_{\mu}W^+_{\nu}
\left(
\partial^{\mu}V^{\nu}
+
\partial^{\nu}V^{\mu}
\right)$ &
dim=4
\\
\hline
$\widetilde{\O}^{W^{+*}W^{-*}V^*}_{2}$ &
$W^-_{\mu}W^+_{\nu}\widetilde{V}^{\mu\nu}$ &
dim=4
\\
\hline
$\widetilde{\O}^{W^{+*}W^{-*}V^*}_{3}$ &
$W^-_{\sigma\mu}W^{+\mu}_{\phantom{+\mu}\nu}
\widetilde{V}^{\nu\sigma}$ &
dim=6
\\
\hline
$\widetilde{\O}^{W^{+*}W^{-*}V^*}_{4}$ &
$\left[
W^-_{\nu}(\partial_{\mu}W^{+\mu})
+
W^+_{\nu}(\partial_{\mu}W^{-\mu})
\right]V^{\nu}$ &
dim=4
\\
\hline
$\widetilde{\O}^{W^{+*}W^{-*}V^*}_{5}$ &
$W^-_{\mu}W^{+\mu}(\partial_{\nu}V^{\nu})$ &
dim=4
\\
\hline
\end{tabular}
\caption{The CP-violating operators for the fully off-shell
$W^{+*}W^{-*}V^*$ vertex, with $V=Z,\gamma$. The first three
structures correspond to the conventional $g_4^V$,
$\widetilde{\kappa}_V$, and $\widetilde{\lambda}_V$ couplings,
respectively, while the last two contain the divergence of a gauge field
and contribute only when the corresponding leg is off shell and
coupled to a non-conserved current.}
\label{tab:WWV-CPv-op}
\end{table}
\end{footnotesize}

%%%%%%%%%%%%%%%%%%%%%%%%%%%%%%%%%%%%
%%%%%%%%%%%%%%%%%%%%%%%%%%%%%%%%%%%%
\end{appendices}
%%%%%%%%%%%%%%%%%%%%%%%%%%%%%%%%%%%%
%%%%%%%%%%%%%%%%%%%%%%%%%%%%%%%%%%%%

%%%%%%%%%%%%%%%%%%%%%%%%%%%%%%%%%%%%
%%%%%%%%%%%%%%%%%%%%%%%%%%%%%%%%%%%%
\bibliographystyle{unsrt}
\bibliography{draft_CV_CPV_Final.bib}

\begin{thebibliography}{10}

\bibitem{Sakharov:1967dj}
A.~D. Sakharov.
\newblock {Violation of CP Invariance, C asymmetry, and baryon asymmetry of the
  universe}.
\newblock {\em Pisma Zh. Eksp. Teor. Fiz.}, 5:32--35, 1967.

\bibitem{Gavela:1993ts}
M.~B. Gavela, P.~Hernandez, J.~Orloff, and O.~Pene.
\newblock {Standard model CP violation and baryon asymmetry}.
\newblock {\em Mod. Phys. Lett. A}, 9:795--810, 1994.

\bibitem{Gavela:1994dt}
M.~B. Gavela, P.~Hernandez, J.~Orloff, O.~Pene, and C.~Quimbay.
\newblock {Standard model CP violation and baryon asymmetry. Part 2: Finite
  temperature}.
\newblock {\em Nucl. Phys. B}, 430:382--426, 1994.

\bibitem{Huet:1994jb}
Patrick Huet and Eric Sather.
\newblock {Electroweak baryogenesis and standard model CP violation}.
\newblock {\em Phys. Rev. D}, 51:379--394, 1995.

\bibitem{Cordero-Cid:2016krd}
A.~Cordero-Cid, J.~Hern{\'a}ndez-S{\'a}nchez, V.~Keus, S.~F. King, S.~Moretti,
  D.~Rojas, and D.~Soko{\l}owska.
\newblock {CP violating scalar Dark Matter}.
\newblock {\em JHEP}, 12:014, 2016.

\bibitem{Cordero-Cid:2018man}
A.~Cordero-Cid, J.~Hern{\'a}ndez-S{\'a}nchez, V.~Keus, S.~Moretti, D.~Rojas,
  and D.~Soko{\l}owska.
\newblock {Lepton collider indirect signatures of dark CP-violation}.
\newblock {\em Eur. Phys. J. C}, 80(2):135, 2020.

\bibitem{Keus:2019szx}
Venus Keus.
\newblock {Dark CP-violation through the $Z$-portal}.
\newblock {\em Phys. Rev. D}, 101(7):073007, 2020.

\bibitem{Keus:2020ooy}
Venus Keus.
\newblock {Dark origins of matter-antimatter asymmetry}.
\newblock {\em PoS}, CORFU2019:059, 2020.

\bibitem{Keus:2021dti}
Venus Keus and Kimmo Tuominen.
\newblock {CP-violating inflation and its cosmological imprints}.
\newblock {\em Phys. Rev. D}, 104(6):063533, 2021.

\bibitem{Keus:2021ojv}
Venus Keus.
\newblock {Baryogenesis from a CP-Violating Inflation}.
\newblock In {\em {Beyond Standard Model: From Theory to Experiment}}, 5 2021.

\bibitem{Keus:2021nff}
Venus Keus.
\newblock {CP-violation in the dark sector}.
\newblock In {\em {55th Rencontres de Moriond on QCD and High Energy
  Interactions}}, 5 2021.

\bibitem{Keus:2024khd}
Venus Keus and Edward~W. Kolb.
\newblock {Baryogenesis from primordial CP violation}.
\newblock {\em JHEP}, 07:156, 2025.

\bibitem{Keus:2025ova}
Venus Keus, Lucy Lewitt, and Jasmine Thomson-Cooke.
\newblock {Electroweak phase transition enhanced by a CP-violating dark
  sector}.
\newblock {\em JHEP}, 09:175, 2026.

\bibitem{Banno:2026hsc}
Tatsuya Banno, Junji Hisano, Teppei Kitahara, Kiyoto Ogawa, and Naohiro
  Osamura.
\newblock {Full three-loop electroweak multiplet contributions to the electron
  electric dipole moment}.
\newblock {\em JHEP}, 05:277, 2026.

\bibitem{Azevedo:2018fmj}
Duarte Azevedo, Pedro~M. Ferreira, M.~Margarete Muhlleitner, Shruti Patel, Rui
  Santos, and Jonas Wittbrodt.
\newblock {CP in the dark}.
\newblock {\em JHEP}, 11:091, 2018.

\bibitem{Abel:2020pzs}
C.~Abel et~al.
\newblock {Measurement of the Permanent Electric Dipole Moment of the Neutron}.
\newblock {\em Phys. Rev. Lett.}, 124(8):081803, 2020.

\bibitem{ACME:2018yjb}
V.~Andreev et~al.
\newblock {Improved limit on the electric dipole moment of the electron}.
\newblock {\em Nature}, 562(7727):355--360, 2018.

\bibitem{Roussy:2022cmp}
Tanya~S. Roussy et~al.
\newblock {An improved bound on the electron{\textquoteright}s electric dipole
  moment}.
\newblock {\em Science}, 381(6653):adg4084, 2023.

\bibitem{Hutzler:2020lmj}
N.~R. Hutzler et~al.
\newblock {Searches for new sources of CP violation using molecules as quantum
  sensors}.
\newblock 10 2020.

\bibitem{Engel:2013lsa}
Jonathan Engel, Michael~J. Ramsey-Musolf, and U.~van Kolck.
\newblock {Electric Dipole Moments of Nucleons, Nuclei, and Atoms: The Standard
  Model and Beyond}.
\newblock {\em Prog. Part. Nucl. Phys.}, 71:21--74, 2013.

\bibitem{CPEDM:2019nwp}
F.~Abusaif et~al.
\newblock {\em {Storage ring to search for electricdipole moments of charged
  particles: Feasibility study}}.
\newblock CERN, Geneva, 6 2021.

\bibitem{Muong-2:2008ebm}
G.~W. Bennett et~al.
\newblock {An Improved Limit on the Muon Electric Dipole Moment}.
\newblock {\em Phys. Rev. D}, 80:052008, 2009.

\bibitem{Pospelov:2005pr}
Maxim Pospelov and Adam Ritz.
\newblock {Electric dipole moments as probes of new physics}.
\newblock {\em Annals Phys.}, 318:119--169, 2005.

\bibitem{Haber:2022gsn}
Howard~E. Haber, Venus Keus, and Rui Santos.
\newblock {P-even, CP-violating signals in scalar-mediated processes}.
\newblock {\em Phys. Rev. D}, 106(9):095038, 2022.

\bibitem{Gaemers:1978hg}
K.~J.~F. Gaemers and G.~J. Gounaris.
\newblock {Polarization Amplitudes for e+ e- ---{\ensuremath{>}} W+ W- and e+
  e- ---{\ensuremath{>}} Z Z}.
\newblock {\em Z. Phys. C}, 1:259, 1979.

\bibitem{Renard:1981es}
F.~M. Renard.
\newblock {Tests of Neutral Gauge Boson Selfcouplings With $e^+ e^- \to \gamma
  Z$}.
\newblock {\em Nucl. Phys. B}, 196:93--108, 1982.

\bibitem{Hagiwara:1986vm}
Kaoru Hagiwara, R.~D. Peccei, D.~Zeppenfeld, and K.~Hikasa.
\newblock {Probing the Weak Boson Sector in e+ e- ---{\ensuremath{>}} W+ W-}.
\newblock {\em Nucl. Phys. B}, 282:253--307, 1987.

\bibitem{Baur:1992cd}
U.~Baur and Edmond~L. Berger.
\newblock {Probing the weak boson sector in $Z \gamma$ production at hadron
  colliders}.
\newblock {\em Phys. Rev. D}, 47:4889--4904, 1993.

\bibitem{Hagiwara:1994pw}
Kaoru Hagiwara, S.~Matsumoto, D.~Haidt, and C.~S. Kim.
\newblock {A Novel approach to confront electroweak data and theory}.
\newblock {\em Z. Phys. C}, 64:559--620, 1994.
\newblock [Erratum: Z.Phys.C 68, 352 (1995)].

\bibitem{Wudka:1994ny}
Jose Wudka.
\newblock {Electroweak effective Lagrangians}.
\newblock {\em Int. J. Mod. Phys. A}, 9:2301--2362, 1994.

\bibitem{Aihara:1995iq}
H.~Aihara et~al.
\newblock {\em {Anomalous gauge boson interactions}}, pages 488--546.
\newblock 3 1995.

\bibitem{CDF:1994bpj}
F.~Abe et~al.
\newblock {Measurement of $W - \gamma$ couplings with {CDF} in $p\bar{p}$
  collisions at $\sqrt{s} = 1.8$ TeV}.
\newblock {\em Phys. Rev. Lett.}, 74:1936--1940, 1995.

\bibitem{Ellison:1998uy}
John Ellison and Jose Wudka.
\newblock {Study of trilinear gauge boson couplings at the Tevatron collider}.
\newblock {\em Ann. Rev. Nucl. Part. Sci.}, 48:33--80, 1998.

\bibitem{Gounaris:1999kf}
G.~J. Gounaris, J.~Layssac, and F.~M. Renard.
\newblock {Signatures of the anomalous $Z_{\gamma}$ and $Z Z$ production at the
  lepton and hadron colliders}.
\newblock {\em Phys. Rev. D}, 61:073013, 2000.

\bibitem{Gounaris:2000dn}
G.~J. Gounaris, J.~Layssac, and F.~M. Renard.
\newblock {Off-shell structure of the anomalous $Z$ and $\gamma$
  selfcouplings}.
\newblock {\em Phys. Rev. D}, 62:073012, 2000.

\bibitem{Buchmuller:1985jz}
W.~Buchmuller and D.~Wyler.
\newblock {Effective Lagrangian Analysis of New Interactions and Flavor
  Conservation}.
\newblock {\em Nucl. Phys. B}, 268:621--653, 1986.

\bibitem{Grzadkowski:2010es}
B.~Grzadkowski, M.~Iskrzynski, M.~Misiak, and J.~Rosiek.
\newblock {Dimension-Six Terms in the Standard Model Lagrangian}.
\newblock {\em JHEP}, 10:085, 2010.

\bibitem{Gounaris:2000tb}
G.~J. Gounaris, J.~Layssac, and F.~M. Renard.
\newblock {New and standard physics contributions to anomalous Z and gamma
  selfcouplings}.
\newblock {\em Phys. Rev. D}, 62:073013, 2000.

\bibitem{Gavela:2014vra}
M.~B. Gavela, J.~Gonzalez-Fraile, M.~C. Gonzalez-Garcia, L.~Merlo, S.~Rigolin,
  and J.~Yepes.
\newblock {CP violation with a dynamical Higgs}.
\newblock {\em JHEP}, 10:044, 2014.

\bibitem{Conti:1992xn}
R.~S. Conti and I.~B. Khriplovich.
\newblock {New limits on T odd, P even interactions}.
\newblock {\em Phys. Rev. Lett.}, 68:3262--3265, 1992.

\bibitem{Ramsey-Musolf:1999cub}
M.~J. Ramsey-Musolf.
\newblock {Electric dipole moments and the mass scale of new T violating, P
  conserving interactions}.
\newblock {\em Phys. Rev. Lett.}, 83:3997--4000, 1999.
\newblock [Erratum: Phys.Rev.Lett. 84, 5681 (2000)].

\bibitem{Kurylov:2000ub}
A.~Kurylov, G.~C. McLaughlin, and M.~J. Ramsey-Musolf.
\newblock {Constraints on T odd, P even interactions from electric dipole
  moments, revisited}.
\newblock {\em Phys. Rev. D}, 63:076007, 2001.

\bibitem{ParticleDataGroup:2024cfk}
S.~Navas et~al.
\newblock {Review of particle physics}.
\newblock {\em Phys. Rev. D}, 110(3):030001, 2024.

\bibitem{Jasinska:2016qsf}
B.~Jasi{\'n}ska et~al.
\newblock {Determination of the 3{\textbackslash}gamma fraction from positron
  annihilation in mesoporous materials for symmetry violation experiment with
  J-PET scanner}.
\newblock {\em Acta Phys. Polon. B}, 47:453, 2016.

\bibitem{Nowakowski:2004cv}
Marek Nowakowski, E.~A. Paschos, and J.~M. Rodriguez.
\newblock {All electromagnetic form-factors}.
\newblock {\em Eur. J. Phys.}, 26:545--560, 2005.

\bibitem{Giunti:2014ixa}
Carlo Giunti and Alexander Studenikin.
\newblock {Neutrino electromagnetic interactions: a window to new physics}.
\newblock {\em Rev. Mod. Phys.}, 87:531, 2015.

\bibitem{Kayser:1982br}
Boris Kayser.
\newblock {Majorana Neutrinos and their Electromagnetic Properties}.
\newblock {\em Phys. Rev. D}, 26:1662, 1982.

\bibitem{Kayser:1984ge}
Boris Kayser.
\newblock {CPT, CP, and C Phases and Their Effects in Majorana Particle
  Processes}.
\newblock {\em Phys. Rev. D}, 30:1023, 1984.

\bibitem{Ivanov:2011ae}
Igor~P. Ivanov, Venus Keus, and Evgeny Vdovin.
\newblock {Abelian symmetries in multi-Higgs-doublet models}.
\newblock {\em J. Phys. A}, 45:215201, 2012.

\bibitem{Ivanov:2012hc}
I.~P. Ivanov and V.~Keus.
\newblock {$Z_p$ scalar dark matter from multi-Higgs-doublet models}.
\newblock {\em Phys. Rev. D}, 86:016004, 2012.

\bibitem{Lavoura:1994fv}
L.~Lavoura and Joao~P. Silva.
\newblock {Fundamental CP violating quantities in a SU(2) x U(1) model with
  many Higgs doublets}.
\newblock {\em Phys. Rev. D}, 50:4619--4624, 1994.

\bibitem{Botella:1994cs}
F.~J. Botella and Joao~P. Silva.
\newblock {Jarlskog - like invariants for theories with scalars and fermions}.
\newblock {\em Phys. Rev. D}, 51:3870--3875, 1995.

\bibitem{Hahn:2010zi}
Thomas Hahn.
\newblock {Feynman Diagram Calculations with FeynArts, FormCalc, and
  LoopTools}.
\newblock {\em PoS}, ACAT2010:078, 2010.

\bibitem{Cordero-Cid:2020yba}
A.~Cordero-Cid, J.~Hern{\'a}ndez-S{\'a}nchez, V.~Keus, S.~Moretti,
  D.~Rojas-Ciofalo, and D.~Soko{\l}owska.
\newblock {Collider signatures of dark $CP$-violation}.
\newblock {\em Phys. Rev. D}, 101(9):095023, 2020.

\bibitem{Keus:2015hva}
Venus Keus, Stephen~F. King, Stefano Moretti, and Kei Yagyu.
\newblock {CP Violating Two-Higgs-Doublet Model: Constraints and LHC
  Predictions}.
\newblock {\em JHEP}, 04:048, 2016.

\bibitem{Keus:2017ioh}
Venus Keus, Niko Koivunen, and Kimmo Tuominen.
\newblock {Singlet scalar and 2HDM extensions of the Standard Model:
  CP-violation and constraints from $(g-2)_\mu$ and $e$EDM}.
\newblock {\em JHEP}, 09:059, 2018.

\bibitem{Chun:2015xfx}
Eung~Jin Chun.
\newblock {The muon g{\ensuremath{-}}2 in two-Higgs-doublet models}.
\newblock {\em EPJ Web Conf.}, 118:01006, 2016.

\bibitem{Cheung:2001hz}
King-man Cheung, Chung-Hsien Chou, and Otto C.~W. Kong.
\newblock {Muon anomalous magnetic moment, two Higgs doublet model, and
  supersymmetry}.
\newblock {\em Phys. Rev. D}, 64:111301, 2001.

\bibitem{Harnik:2012pb}
Roni Harnik, Joachim Kopp, and Jure Zupan.
\newblock {Flavor Violating Higgs Decays}.
\newblock {\em JHEP}, 03:026, 2013.

\bibitem{Georgi:1978ri}
Howard Georgi and Dimitri~V. Nanopoulos.
\newblock {Suppression of Flavor Changing Effects From Neutral Spinless Meson
  Exchange in Gauge Theories}.
\newblock {\em Phys. Lett. B}, 82:95--96, 1979.

\bibitem{Lavoura:1994yu}
L.~Lavoura.
\newblock {Signatures of discrete symmetries in the scalar sector}.
\newblock {\em Phys. Rev. D}, 50:7089--7092, 1994.

\bibitem{Branco:1999fs}
Gustavo~C. Branco, Luis Lavoura, and Joao~P. Silva.
\newblock {\em {CP Violation}}, volume 103.
\newblock 1999.

\bibitem{Davidson:2005cw}
Sacha Davidson and Howard~E. Haber.
\newblock {Basis-independent methods for the two-Higgs-doublet model}.
\newblock {\em Phys. Rev. D}, 72:035004, 2005.
\newblock [Erratum: Phys.Rev.D 72, 099902 (2005)].

\bibitem{Ginzburg:2001wj}
Ilya~F. Ginzburg, Maria Krawczyk, and Per Osland.
\newblock {Potential of photon collider in resolving SM like scenarios}.
\newblock {\em Nucl. Instrum. Meth. A}, 472:149--154, 2001.

\bibitem{Gunion:2002zf}
John~F. Gunion and Howard~E. Haber.
\newblock {The CP conserving two Higgs doublet model: The Approach to the
  decoupling limit}.
\newblock {\em Phys. Rev. D}, 67:075019, 2003.

\bibitem{Craig:2013hca}
Nathaniel Craig, Jamison Galloway, and Scott Thomas.
\newblock {Searching for Signs of the Second Higgs Doublet}.
\newblock 5 2013.

\bibitem{Asner:2013psa}
D.~M. Asner et~al.
\newblock {ILC Higgs White Paper}.
\newblock In {\em {Snowmass 2013}: {Snowmass on the Mississippi}}, 10 2013.

\bibitem{Carena:2013ooa}
Marcela Carena, Ian Low, Nausheen~R. Shah, and Carlos E.~M. Wagner.
\newblock {Impersonating the Standard Model Higgs Boson: Alignment without
  Decoupling}.
\newblock {\em JHEP}, 04:015, 2014.

\bibitem{Haber:2013mia}
Howard~E. Haber.
\newblock {The Higgs data and the Decoupling Limit}.
\newblock In {\em {1st Toyama International Workshop on Higgs as a Probe of New
  Physics 2013}}, 12 2013.

\bibitem{Degrande:2013kka}
Celine Degrande.
\newblock {A basis of dimension-eight operators for anomalous neutral triple
  gauge boson interactions}.
\newblock {\em JHEP}, 02:101, 2014.

\bibitem{Subba:2023rpm}
Amir Subba and Ritesh~K. Singh.
\newblock {Study of anomalous $W^-W^+\gamma /Z$ couplings using polarizations
  and spin correlations in $e^-e^+\rightarrow W^-W^+$ with polarized beams}.
\newblock {\em Eur. Phys. J. C}, 83(12):1119, 2023.

\bibitem{Ellis:2023ucy}
John Ellis, Hong-Jian He, and Rui-Qing Xiao.
\newblock {Probing neutral triple gauge couplings with Z*{\ensuremath{\gamma}}
  ({\ensuremath{\nu}}{\ensuremath{\nu}}{\textasciimacron}{\ensuremath{\gamma}})
  production at hadron colliders}.
\newblock {\em Phys. Rev. D}, 108(11):L111704, 2023.

\bibitem{Ellis:2024omd}
John Ellis, Hong-Jian He, Rui-Qing Xiao, Shi-Ping Zeng, and Jiaming Zheng.
\newblock {UV completion of neutral triple gauge couplings}.
\newblock {\em Phys. Rev. D}, 111(1):015007, 2025.

\end{thebibliography}
%%%%%%%%%%%%%%%%%%%%%%%%%%%%%%%%%%%%
%%%%%%%%%%%%%%%%%%%%%%%%%%%%%%%%%%%%

\end{document}

%%%%%%%%%%%%%%%%%%%%%%%%%%%%%%%%%%%%
%%%%%%%%%%%%%%%%%%%%%%%%%%%%%%%%%%%%
%%%%%%%%%%%%%%%%%%%%%%%%%%%%%%%%%%%%
%%%%%%%%%%%%%%%%%%%%%%%%%%%%%%%%%%%%
%%%%%%%%%%%%%%%%%%%%%%%%%%%%%%%%%%%%
%%%%%%%%%%%%%%%%%%%%%%%%%%%%%%%%%%%%